\documentclass[letterpaper,twocolumn,english,showpacs, superscriptaddress]{iopart}
\usepackage[T1]{fontenc}
\usepackage[latin9]{inputenc}
\usepackage{babel}
\usepackage{verbatim}
\usepackage{amstext}
\usepackage{amssymb}
\usepackage{graphicx}
\usepackage{esint}
\usepackage{color} 
\usepackage{hyperref}
\usepackage{breakurl}
\usepackage{soul}
\makeatletter

\usepackage{iopams}
\usepackage{setstack}

\usepackage{babel}

\usepackage[numbers, sort&compress]{natbib}

\def\newblock{\hskip .11em plus .33em minus .07em}

\makeatother
\begin{document}

\title{Tunneling transport in NSN Majorana junctions  across
the topological quantum phase transition}

\author{Alejandro M. Lobos}

\address{Facultad de Ciencias Exactas Ingenier\'ia y Agrimensura, Universidad
Nacional de Rosario and Instituto de F\'isica Rosario,  Bv. 27 de Febrero
210 bis, 2000 Rosario, Argentina}

\address{}

\address{Condensed Matter Theory Center and Joint Quantum Institute, Department
of Physics, University of Maryland, College Park, Maryland 20742-4111,
USA.}

\ead{lobos@ifir-conicet.gov.ar}
	
\author{S. Das Sarma}

\address{Condensed Matter Theory Center and Joint Quantum Institute, Department
of Physics, University of Maryland, College Park, Maryland 20742-4111,
USA.}

\date{\today}
\begin{abstract}
We theoretically consider transport properties of a normal metal (N)-
superconducting semiconductor nanowire (S)-normal metal (N) structure
(NSN) in the context of the possible existence of Majorana bound states
in semiconductor-superconductor hybrid systems with
spin-orbit coupling and external magnetic
field. We study in detail the transport signatures of
the topological quantum phase transition as well as the existence
of the Majorana bound states in the electrical transport properties
of the NSN structure. Our treatment includes the realistic non-perturbative
effects of disorder, which is detrimental to the topological phase
(eventually suppressing the superconducting gap completely), and the
effects of the tunneling barriers (or the transparency at the tunneling
NS contacts), which affect (and suppress) the zero bias conductance
peak associated with the zero-energy Majorana bound states. We show
that in the presence of generic disorder and barrier transparency
the interpretation of the zero bias peak as being associated with
the Majorana bound state is problematic since the non-local correlations
between the two NS contacts at two ends may not manifest themselves
in the tunneling conductance through the whole NSN structure. We establish
that a simple modification of the standard transport measurements
using conductance differences (rather than the conductance itself
as in a single NS junction) as the measured quantity can allow direct
observation of the non-local correlations inherent in the Majorana
bound states. We also show that our proposed analysis of transport
properties of the NSN junction enables the mapping out of the topological
phase diagram (even in the presence of considerable disorder) by precisely
detecting the topological quantum phase transition point. We propose
direct experimental studies of NSN junctions (rather than just a single
NS junction) in order to establish the existence of Majorana bound
states and the topological superconducting phase in semiconductor
nanowires of current interest. Throughout the work we emphasize that
the NSN transport properties are sensitive to both the bulk topological
phase and the end Majorana bound states, and thus the NSN junction
is well-suited for studying the non-local correlations between the
end Majorana modes as well as the bulk topological quantum phase transition
itself.
\end{abstract}


\maketitle

\section{\label{sec:intro}Introduction}

The subject of topological superconductors (SCs) hosting non-Abelian
quasiparticles has become one of the most intensively investigated
topics in condensed matter physics.\cite{Read00_Topological_SC_in_2D,kitaev2001}
In particular, one-dimensional topological superconductors have been
predicted to support zero-energy particle-hole symmetric non-Abelian
Majorana bound-states (MBS) localized at the ends.\cite{kitaev2001}
Beyond their intrinsic fundamental interest, MBS have attracted attention
for their potential use in fault-tolerant topological quantum computation
schemes.\cite{Nayak08} Far from being a subject of purely theoretical
interest, concrete experimental proposals to realize these exotic
states of matter have been put forward recently,\cite{Fu08,Sau10_Proposal_for_MF_in_semiconductor_heterojunction,Lutchyn'10,Oreg'10,Sau10_long}
some of which have been implemented experimentally.\cite{Mourik12_Signatures_of_MF,Das12_Evidence_of_MFs,Deng12_ZBP_in_Majorana_NW,Rokhinson2012,Finck13_ZBP_in_hybrid_NW_SC_device,Churchill2013}
In particular, Refs. \cite{Lutchyn'10,Oreg'10,Sau10_long} showed
that a one-dimensional semiconductor nanowire in proximity to a bulk
s-wave superconductor, and subjected to strong Rashba spin-orbit coupling
can be driven into a topologically non-trivial phase with MBS localized
at the ends, upon the application of an external Zeeman magnetic field.
In this topologically non-trivial phase, the nanowire becomes \textit{effectively}
a helical spinless p-wave superconductor, realizing an idea originally
proposed by Kitaev for the localization of isolated MBS in a physical
system.\cite{kitaev2001} Other experimental setups involving arrays
of magnetic atoms on s-wave SCs,\cite{Nadj-Perdge13_Majorana_fermions_in_Shiba_chains}
or cold-atomic systems \cite{Jiang11_MFs_in_Cold_Atoms} have also
been proposed and are currently under experimental consideration.
It is important to mention here that the real significance of course
is the creation of isolated zero-energy MBS at the ends of the nanowire
which are well-separated from each other so that they can be considered
topologically protected.

On the experimental side, one of the most relevant questions is how
to establish the presence of ``true'' MBS in a real experiment.
In principle, the tunneling conductance at the end of the topological
SC nanowire should reveal an MBS as a quantized zero-bias peak (ZBP)
of magnitude $2e^{2}/h$ in the conductance at zero temperature, which
is a direct manifestation of the perfect Andreev reflection associated
with the MBS.\cite{Sengupta01_Midgap_states_in_1D_conductors,Law09,Sau10_long,Flensberg10_Quantization_MBS,Prada12_Transport_through_NS_junctions_with_MBS,Roy12_MF_out_of_equilibrium_transport}
Recent experiments implementing the proposal in Refs. \cite{Lutchyn'10, Oreg'10,Sau10_long}
have shown an intriguing ZBP, in apparent agreement with theoretical
predictions for the existence of MBS, which appears upon application
of a Zeeman field, providing compelling preliminary evidence of the
Majorana scenario.\cite{Mourik12_Signatures_of_MF,Das12_Evidence_of_MFs,Deng12_ZBP_in_Majorana_NW}
However, the interpretation of these experiments seems to be considerably
more complex than the ideal models originally proposed and show several
deviations from the predicted behavior, among which we mention the
most important ones: a) the smallness of the ZBP in comparison to the ideally
theoretical value of $2e^{2}/h$ (i.e., $0.1-0.2\ e^{2}/h$ in the
low temperature limit), b) the presence of a continuum of fermionic
excitations in the subgap region (i.e., the so-called ``soft-gap''
feature) instead of a well-defined SC gap, and c) the lack of evidence
for the closing and then reopening of this SC soft-gap upon increasing
the Zeeman field across the putative critical field $V_{c}$. We stress
that this last feature is a crucial prerequisite for the existence
of MBS, which would be indicative of a topological quantum phase transition
(TQPT) taking place in the sample where the gap must vanish.%
{} 

Contrasting with the interpretation that the recent nanowire experimental
observations are indeed evidence for the isolated existence of MBS
in a topological SC system,%
{} it has been pointed out that other ZBPs (or near-ZBPs) sharing similar
features with the MBS are generically allowed in spin-orbit-coupled
nanowires subject to a magnetic field in the presence of disorder
or smooth confining potentials, both in the topologically trivial
and non-trivial phases, a fact that would hinder the observation of
bona fide Majorana-type excitations.\cite{Pikulin12_ZBP_from_weak_antilocalization_in_Majorana_NW,Liu12_ZBP_in_Majorana_wires_with_and_without_MZBSs,Bagrets12_Class_D_spectral_peak_in_Majorana_NW,Kells12_Near-zero-end_states_in_Majorana_wires_with_smooth_confinement,Rieder12_Endstates_in_multichannel_p-wave_SC_nanowires}
In particular, disorder is known to have strong detrimental effects
in p-wave SCs.\cite{Motrunich01_Disorder_in_topological_1D_SC,Brouwer00_Localization_Dirty_SC_wire,Gruzberg05_Localization_in_disordered_SC_wires_with_broken_SU2_symmetry,Brouwer11_Probability_distribution_of_MFS_in_disordered_wires,Brouwer11_Topological_SC_in_disorder_wires,Lobos12_Interplay_disorder_interaction_Majorana_wire,Akhmerov11_Quantized_conductance_in_disordered_wire,DeGottardi11_MFs_with_disorder,DeGottardi_MFs_with_spatially_varying_potentials,Sau12_MF_in_real_materials,Lin12_ZBP_in_SM_SC_structures,Adagideli13_Topological_order_in_dirty_wires,Fregoso13_Electrical_detection_of_TQPT,Sau13_DOS_of_disordered_Majorana_wire,Rainis2013,Hui14_Eilenberger_theory_for_MBS}
Motrunich \textit{et al.} showed more than a decade ago that Andreev
subgap states induced by disorder tend to proliferate in one-dimensional
systems described by Bogoliubov-de Gennes Hamiltonians with broken
time and spin-rotational symmetry (symmetry class D, like the nanowires
in Refs. \cite{Lutchyn'10, Oreg'10, Sau10_long}), and render
the system gapless.\cite{Motrunich01_Disorder_in_topological_1D_SC}
These authors predicted that for weak disorder an infinite system
realizes a topologically non-trivial phase with two degenerate zero-energy
MBS localized at the ends of the wire. In a finite-length system of
size $L_{\text{w}}$, this degeneracy is lifted by an exponential
splitting $\Delta\varepsilon\sim e^{-L_{\text{w}}/\xi}$, where $\xi$
is the superconducting coherence length. Increasing the amount of
disorder generates low-energy Andreev bound states, and the (averaged)
scaling of the splitting energy changes to $\Delta\varepsilon\sim e^{-L_{\text{w}}/\xi+L_{\text{w}}/(2\ell_{e})}$,
where $\ell_{e}$ is the elastic mean-free path of the system.\cite{Brouwer11_Probability_distribution_of_MFS_in_disordered_wires}
Beyond a critical disorder amount, defined by the condition $\ell_{e}=\xi/2$,
the system experiences a TQPT induced by disorder and enters a non-topological
insulating phase with no end-MBS. At both sides of the TQPT, the system
is \textit{localized} at zero energy, and exactly at the critical
point separating these phases, the wave functions become \textit{delocalized}
and the smallest Lyapunov exponent (i.e., the inverse of the localization
length of the system) vanishes. This intimate connection between localization
and topology in disordered topological superconductors has been stressed
 in a series of theoretical works.\cite{Motrunich01_Disorder_in_topological_1D_SC,Akhmerov11_Quantized_conductance_in_disordered_wire,DeGottardi11_MFs_with_disorder,Fregoso13_Electrical_detection_of_TQPT}
The interplay among disorder, superconductivity, and possible Majorana
zero modes is still very much an important open problem in the subject,
and whether the experimentally observed ZBP is indeed the manifestation
of the theoretically predicted MBS can only be sorted out definitively
by accurately understanding the precise role of disorder in the experimental
systems. In particular, a key question is the effect of disorder on
the TQPT itself, which is a central topic of the current work.

Concerning the rather ubiquitous presence of in-gap states (or ``soft
gap'') in the experiment, it is important to note the lack of evidence
of a well-defined superconducting gap in most of the experiments involving evaporated SC-semiconductor SN contacts, even in absence of an applied
magnetic field, when the time-reversal symmetry is not broken (i.e.,
symmetry class DIII or BDI). 
By improving the quality of the semiconductor/SC interface using molecular
beam-epitaxy methods, as was theoretically predicted \cite{Takei13_Soft_gap}, recent experiments have reported much harder
gaps \cite{Ziino13_Hard_gap_by_MBE, Chang15_Hard_gap_in_SM_SC_NWs}, suggesting that some sort of disorder at the might be operative at the SN interface.
Since the topological protection
of the MBS is directly provided by the existence of the SC gap, it
is of obvious importance to understand the physical origin of this soft gap for the correct  interpretation of the experiment (as well as to help produce hard gap
systems for future Majorana experiments).  
Stanescu \textit{et al.} have suggested recently that intrinsic quasiparticle
broadening effects due to the hybridization of the SC with the normal
metallic lead, could explain this feature.\cite{Stanescu13_SoftGap}
Indeed, it is well-known that a highly transparent NS barrier can
produce large subgap conductance\cite{Blonder1982_BTK_paper} {[}i.e.
Blonder-Tinkham-Klapwijk (or BTK) barrier parameter $Z\rightarrow0${]},
and therefore could also induce a large broadening of Bogoliubov quasiparticles,
and hence a soft gap through this ``inverse proximity effect'' of
the normal metallic lead on the SC nanowire. However, this does not
seem to be the complete explanation of the experiment. Recent experiments
where the transparency of the NS contact was systematically reduced
have shown that the soft gap persists even in the low-transparency
limit (i.e., ``pinching off'' the quantum point contact when the
inverse proximity effect should be exponentially suppressed).\cite{Churchill2013, Chang15_Hard_gap_in_SM_SC_NWs}
An alternative explanation for the soft gap, valid in the limit of
low transparency (i.e., large BTK barrier parameter $Z\rightarrow\infty$),
was proposed in Ref. \cite{Takei13_Soft_gap}. Among the many
different pair-breaking mechanisms that might be operative in Majorana
nanowires as considered in Ref. \cite{Takei13_Soft_gap} (e.g.,
finite temperature $T$, presence of magnetic impurities, quasiparticle
broadening, etc.) realistic parameter considerations point to the
predominance of a special kind of inhomogeneity, which was not considered
before in the present context: the spatial fluctuations in the proximity-induced
pair potential $\Delta\left(x\right)$.\cite{Takei13_Soft_gap} Physically,
spatial fluctuations in $\Delta\left(x\right)$ are likely to be introduced
by disorder or inhomogeneities at the SC/semiconductor contact.\cite{note_majorana_long_paper}
Following the suggestion of the
theoretical explanation given in Ref. \cite{Takei13_Soft_gap}, the above-mentioned experiments involving epitaxially grown SC/semiconductor nanowires \cite{Ziino13_Hard_gap_by_MBE, Chang15_Hard_gap_in_SM_SC_NWs} have reported much harder gaps.  
This constitutes a qualitative improvement in the fabrication of Majorana
nanowires, and hopefully a new generation of experiments where disorder
effects are dramatically reduced will be soon available with hard
proximity gaps (i.e. no subgap fermionic excitations) and well-defined
MBS. We incorporate this aspect of the soft gap physics in the current
work through a simple model approximation which mimics the spatial
variation in the proximity-induced superconducting pair potential
arising from the inhomogeneities at the superconductor-nanowire interface
{[}see Eq. (\ref{eq:Delta_profile}) below and the associated discussion{]}.

The above discussion describes the rather complex situation faced
in the experiments in order to detect ``true'' MBS in the topological
phase. In this article we focus on a specific configuration, the normal-topological
superconductor-normal (NSN) configuration, which is currently under
experimental study. The SC part of this NSN (i.e. the S-part) junction
is the semiconductor nanowire which has proximity-induced superconductivity
from an underlying ordinary s-wave SC system. Many of the recent experiments
have focused specifically on just the simple NS junction, but NSN
junctions are essentially ``equally easy'' to study, and they have
been studied also. We believe that NSN junctions have some intrinsic
advantages over the minimal NS junction transport for studying MBS
physics and the associated TQPT. We provide a comprehensive theoretical
analysis of its transport properties taking into account the effects
of disorder, inhomogeneities and temperature. As noted in previous
works, the NSN configuration allows to extract the same information
as in the simpler NS contacts, but contains additional interesting
new physics arising from non-local correlations.\cite{Bolech07_Shot_noise_Majorana,Nilsson_PRL08,Liu12_Current_noise_correlation,Zocher13_Current_cross_correlations_in_Majorana_NW} 

The current work is a generalization and extension of our earlier
work in Ref. \cite{Fregoso13_Electrical_detection_of_TQPT},
where we introduced an original proposal for a direct experimental
study of the Majorana fermion-related TQPT in hybrid semiconductor
nanowire structures. Here, we present a more detailed study of the
tunneling transport properties of the NSN junction, a fact that allows
us to make contact with recent and ongoing experiments. \cite{Mourik12_Signatures_of_MF,Das12_Evidence_of_MFs,Deng12_ZBP_in_Majorana_NW,Finck13_ZBP_in_hybrid_NW_SC_device,Churchill2013}
In contrast to our previous Ref. \cite{Fregoso13_Electrical_detection_of_TQPT},
where we computed the differential conductance \textit{only} at one
end of the NSN system, at zero bias voltage and at zero temperature,
in this work we extend our calculation to the \textit{full} differential
conductance matrix (see Eq. \ref{eq:Gmatrix}) at finite bias voltage.
In addition, we also study the thermal effects (see Fig. \ref{fig:dGLL}),
which are important in order both to quantify the detrimental effects
on the efficency of our proposed detection scheme for the TQPT {[}see
Eq. (\ref{eq:dGjj}){]}, as well as for allowing a more realistic
comparison with the experiments. Finally, in this work we also provide
a physically intuitive theoretical description (see Section \ref{sec:intuitive})
of the proposed experiment in terms of an exaclty solvable ``random-mass''
Dirac model, where the interplay between disorder, external magnetic
fields, and the emergence or destruction of MBS, is made fully transparent. 

While this is not a ``smoking-gun'' experiment, it might be an extremely
useful experimental tool providing information about the topological
phase diagram of the system, complementary to non-local shot noise
correlations.\cite{Bolech07_Shot_noise_Majorana,Nilsson_PRL08,Liu12_Current_noise_correlation,Zocher13_Current_cross_correlations_in_Majorana_NW}
Observation of non-local correlations as well as studying the TQPT
itself using our suggested transport techniques in NSN junctions taken
together may in fact serve as the smoking gun evidence for the existence
of Majorana modes in nanowire systems. An associated significant advantage
of the NSN junctions over the much-studied NS junctions in the context
of Majorana physics in nanowires, which should be obvious from the
above discussion and is emphasized throughout this work, is that transport
in NSN junctions potentially studies both the end MBS and the bulk
topological SC phase whereas NS junction tunneling properties may
very well be dominated by the end MBS so as to suppress the manifestation
of the bulk TQPT and the non-local correlations between the two end
MBS which must go through the bulk nanowire. This is the key reason
for our promoting NSN junction transport studies as an important tool
for the Majorana investigation.

\begin{figure}[t]
\begin{centering}
\includegraphics[bb=30bp 40bp 700bp 550bp,clip,scale=0.35]{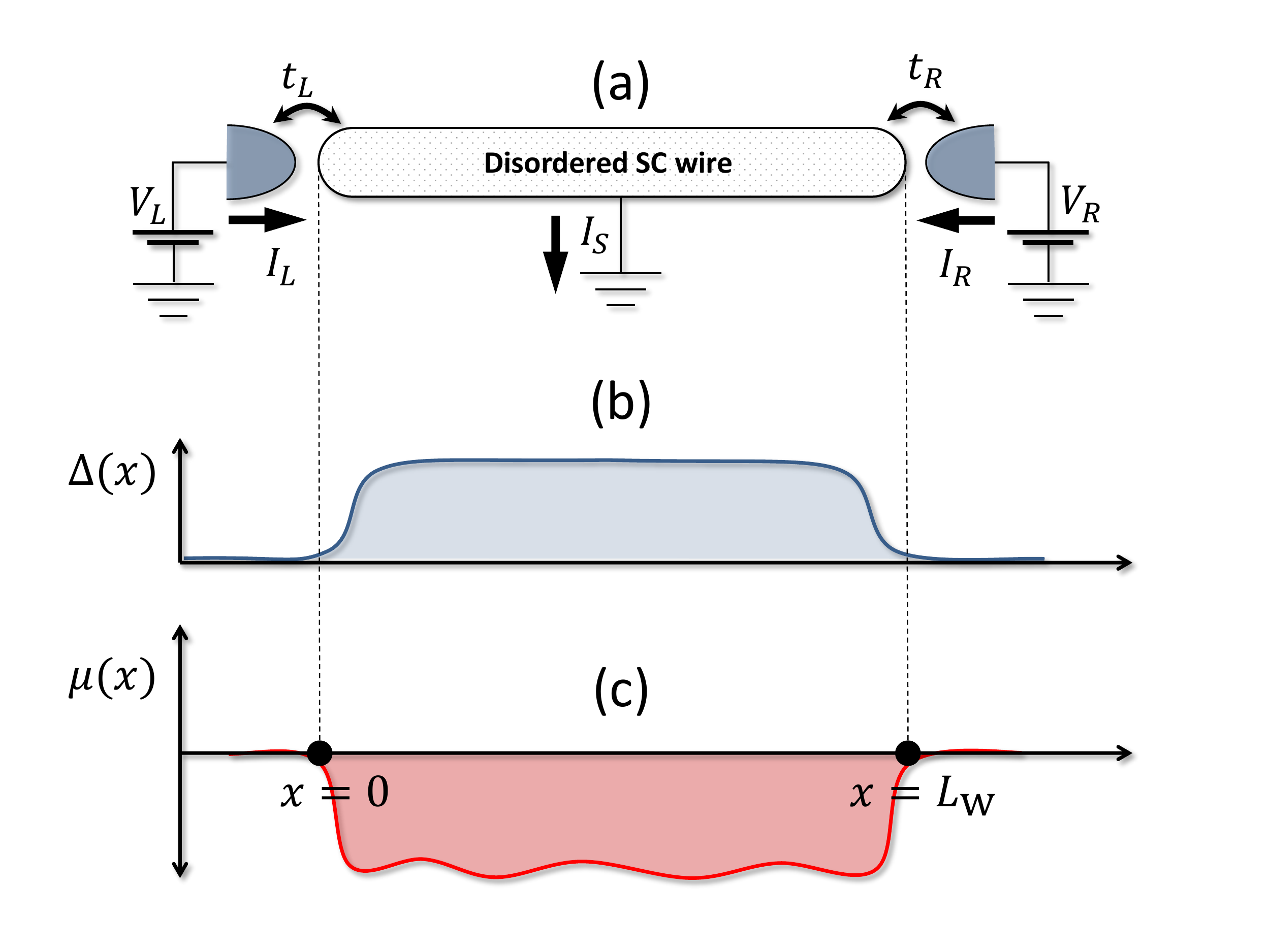}
\caption{(a) Schematic diagram of a NSN circuit where the superconducting (S)
part corresponds to the proximity-induced semiconductor Majorana nanowire.
Figures (b) and (c) correspond to the proximity-induced superconducting
pair potential $\Delta\left(x\right)$ and chemical potential $\mu\left(x\right)$
profiles, respectively.\label{fig:disordered_wire}}
\end{centering}\end{figure}

\begin{figure}[t]
\begin{centering}
\includegraphics[bb=0bp 270bp 620bp 550bp,clip,scale=0.6]{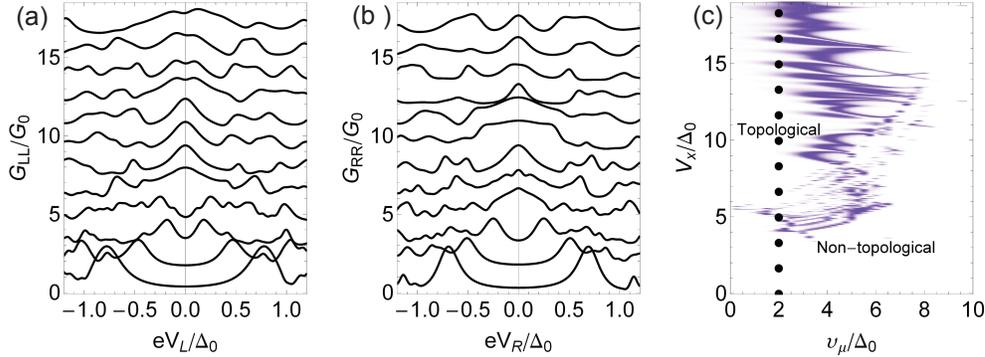}\caption{\label{fig:GLL_GRR_pdiag}(a) Differential conductance at the left
end $G_{LL}$ as a function of the left bias voltage $eV_{L}$, for
different values of the Zeeman field $V_{x}$, and at a temperature
$T/\Delta_{0}=0.02$, which corresponds to the experimental temperature
$T_{\text{exp}}\approx60$ mK (see e.g. Ref. \cite{Mourik12_Signatures_of_MF}).
The curves have been shifted vertically for clarity. (b) idem for
the conductance at the right end $G_{RR}$ as a function of the right
bias voltage $eV_{R}$. In both Figs. \ref{fig:GLL_GRR_pdiag}(a)
and \ref{fig:GLL_GRR_pdiag}(b) the zero-bias peaks are smeared by
temperature, disorder and quasiparticle broadening arising from the
coupling to the normal leads, and they appear at different values
of the Zeeman field {[}see Figs. \ref{fig:Gall}(a) and \ref{fig:Gall}(b)
below for more details{]}, complicating the physical interpretation.
(c) Color map of the thermal transmission probability $\mathcal{T}_{1N}$
as function of Zeeman field $V_{x}$ and disorder strength $\upsilon_{\mu}$.
The blue regions correspond to values close to the maximum $\mathcal{T}_{1N}=1$
and therefore correspond to the location of the topological quantum
phase transition. Each dot corresponds to each one of the curves in
Figs. \ref{fig:GLL_GRR_pdiag}(a) and \ref{fig:GLL_GRR_pdiag}(b).}
\end{centering}
\end{figure}

In order to illustrate the main motivation of this article, let us
first consider a ``dirty'' proximity-induced SC Majorana nanowire,
electrically connected to ground and attached to normal contacts in
a NSN configuration, as shown schematically in Fig. \ref{fig:disordered_wire}(a).
Here we consider a generic situation where inhomogeneities are present
both in the form of spatial fluctuations of the (proximity-induced)
pairing potential, and in the form of quenched disorder in the on-site
chemical potential fluctuations {[}Figs. \ref{fig:disordered_wire}(b)
and (c){]}. We also assume an external Zeeman field applied in the
direction parallel to the nanowire, which allows to drive the system
across the TQPT. A relevant experimental quantity is the differential
conductance matrix, defined as 

\begin{eqnarray}
G_{ij}\left(eV_{j}\right) & \equiv&\frac{dI_{i}}{dV_{j}}\left(eV_{j}\right),\label{eq:Gmatrix}
\end{eqnarray}
where $I_{i}$ and $V_{j}$ are, respectively, the current and voltage
applied in the $\left\{ i,j\right\} =\left\{ L,R\right\} $ normal
contact. In ideal conditions, the local conductances $G_{LL}$ and
$G_{RR}$ should reveal the presence of end-MBS as a quantized ZBP
peak of magnitude $2e^{2}/h$ at $T=0$.\cite{Sengupta01_Midgap_states_in_1D_conductors,Law09,Sau10_long,Flensberg10_Quantization_MBS}
In practice, however, disorder, finite temperatures, quasiparticle
poisoning, etc., might hinder or even destroy the purported topological
phases and, therefore, the MBS. Since we are motivated by the current
experiments, we start by showing a typical example of our numerical
simulations of tunneling transport in Figs. \ref{fig:GLL_GRR_pdiag}(a)
and \ref{fig:GLL_GRR_pdiag}(b), and leave the explanation of the
theoretical details for Secs. \ref{sec:model},\ref{sec:transfer}
and \ref{sec:transport}. In these plots have computed the local conductances
$G_{LL}$ and $G_{RR}$ for a disordered wire at a finite temperature
as a function of the local bias voltages $V_{L}$ and $V{}_{R}$,
respectively, and for different values of the applied Zeeman field.
In contrast to the ideal case\cite{Stanescu11_MFs_in_SM_nanowires}
(i.e., clean system and $T=0$), where a vanishing single-particle
excitation gap signals the TQPT across the critical Zeeman field,
with the ZBP emerging on the topological side at higher magnetic field,
here the presence of the above mentioned non-idealities renders the
situation much less clear to determine the TQPT and the nature of
the ZBPs. In other words, the information about the ZBP has been ``washed out''
by a combination of thermal effects, disorder and quasiparticle broadening,
although the conductance results in Fig. \ref{fig:GLL_GRR_pdiag}
are explicitly obtained theoretically in a system where the MBS definitively
exists in the ideal situation. (As an aside, we mention that the theoretical
conductance results depicted numerically in Fig. \ref{fig:GLL_GRR_pdiag}
 look remarkably similar to the measured tunneling spectroscopy results
reported so far in the literature in the context of Majorana nanowire
experiments.). The ZBPs emerge in a soft-gap background and, in agreement
with recent experimental results, the left and right ZBPs appear and
disappear at different values of the Zeeman field (i.e., they appear
not to be correlated). Is the wire ``fragmented'', so that the end
Majoranas do not know about the existence of each other? What is the
topological state of the nanowire? Does the wire have more than one
pair of MBS because of disorder? How do we establish the existence
of MBS using such imperfect ZBP data in a manifestly soft gap situation?
These are the kind of questions that motivate our work.

The article is divided as follows. In Sec. \ref{sec:model} we present
the theoretical framework, the model and the main approximations.
In Sec. \ref{sec:transfer} we describe the method used to determine
theoretically the topological phase diagram of a disordered Majorana
wire. In Sec. \ref{sec:transport} we present the theoretical technique
to describe the differential conductance of a generic disordered Majorana
wire in the NSN configuration and analyze the physical content in
the analytical expressions. In Sec. \ref{sec:pdiag} we describe in
detail a proposal to extract information about the TQPT and to assess
the topological stability of MBS. Sec. \ref{sec:intuitive} is intended
to provide a simple intuitive theoretical understanding the physics
underlying our proposal, in Sec. \ref{sec:summary} we present a summary
and our conclusions, and finally in Appendix \ref{sec:appendix} we
give a detailed derivation of the Eqs. (\ref{eq:GLL_final_symm-1})-(\ref{eq:GRR_final_symm-1})
for the conductance matrix in the NSN configuration.

\section{\label{sec:model}Theoretical model}

In accordance with previous works on Majorana wires,\cite{Lutchyn'10,Oreg'10,Stanescu11_MFs_in_SM_nanowires}
we consider the following Hamiltonian describing a disordered semiconductor
nanowire of length $L_{\text{w}}$, subjected to Rashba spin-orbit
coupling and a Zeeman field, $H_{\text{NW}}=H_{0}+H_{\Delta}$, where

\begin{eqnarray}
H_{0} & =&\int_{0}^{L_{\text{w}}}dx\;\psi_{\sigma}^{\dagger}\left(x\right)\biggl[-\frac{\partial_{x}^{2}}{2m}-\mu\left(x\right)+i\alpha_{R}\hat{\boldsymbol{\sigma}}_{y}\partial_{x}\nonumber \\
 && +V_{x}\hat{\boldsymbol{\sigma}}_{x}\biggr]_{\sigma\sigma^{\prime}}\psi_{\sigma^{\prime}}\left(x\right),\label{eq:H_0}\\
H_{\Delta} & =&\int_{0}^{L_{\text{w}}}dx\;\Delta\left(x\right)\left[\psi_{\uparrow}^{\dagger}\left(x\right)\psi_{\downarrow}^{\dagger}\left(x\right)+\psi_{\downarrow}\left(x\right)\psi_{\uparrow}\left(x\right)\right].\label{eq:H_SC}
\end{eqnarray}
Here, $\psi_{\sigma}^{\dagger}\left(x\right)$ creates a fermion with
spin projection $\sigma$, and $\hat{\boldsymbol{\sigma}}_{i}$ (with
$i=x,y,z$) are the Pauli matrices acting on spin space. The parameter
$\alpha_{R}$ is the Rashba spin-orbit coupling strength and $V_{x}$
is the Zeeman field along the wire, and summation over repeated indices
$\sigma$ is implied. The term $H_{\Delta}$ represents the effect
of a proximate bulk s-wave SC on the nanowire {[}not shown in Fig.
\ref{fig:disordered_wire}(a){]}, which induces a mean-field SC pairing
potential $\Delta\left(x\right)$ through the proximity effect. For
simplicity, we have assumed single-channel occupancy in the nanowire
with no loss of generality. As we will explain later, our results
are generic and this single-channel (or single-subband) assumption
does not affect the main conclusions in the case of many occupied
subbands (as long as an odd number of subbands are occupied which
is a necessary condition for the existence of the MBS for many occupied
subbands \cite{Lutchyn2011}). We recall that $H_{\text{NW}}$ is only an effective one-dimensional
model describing the system at low temperatures. A more realistic
model should involve an explicit coupling $t_{\perp}$ to the proximate
bulk SC, which is the source of superconducting correlations, and
a self-consistent determination of $\Delta\left(x\right)$. However,
this task is beyond the scope of this work and does not change our
results qualitatively since all we need in our model is the existence
of a pairing potential in the nanowire. For more details, we refer
the reader to Refs. \cite{Stanescu11_MFs_in_SM_nanowires,BlackSchaffer11_Selfconsistent_proximity_effect_at_SQH_edge}
where a deeper discussion on this issue is provided, which is not
particularly germane for our consideration in the current work where
we are interested in the realistic manifestation of the MBS themselves
rather the issue of proximity effect.

Disorder and inhomogeneities enter in the above model through two
physically different mechanisms: a) Local fluctuations of the chemical
potential $\mu\left(x\right)=\mu_{0}+\delta\mu\left(x\right)$, with
$\mu_{0}$ a uniform value which in principle can be controlled by
external gates, and the fluctuations $\delta\mu\left(x\right)$ are
physically related to the presence of impurities, vacancies, etc.
in the environment (both the nanowire itself and the surrounding).
We assume $\delta\mu\left(x\right)$ to be a Gaussian random variable
fully characterized by $\left\langle \delta\mu\left(x\right)\right\rangle =0$
and $\left\langle \delta\mu\left(x\right)\delta\mu\left(y\right)\right\rangle =\upsilon_{\mu}^{2}\delta\left(x-y\right)$,
with the standard deviation $\upsilon_{\mu}$ representing the ``strength''
of disorder {[}see the horizontal axis in Fig. \ref{fig:GLL_GRR_pdiag}(c){]}.
For one single realization of disorder, once the nanowire is deposited
and electrically contacted, we assume this parameter to be fixed throughout
the experiment. b) Local variations in the (induced) pair potential
$\Delta\left(x\right)$, which for concreteness (and numerical convenience)
here we model as

\begin{eqnarray}
\Delta\left(x\right) & =&\Delta_{0}\tanh\left(\frac{x}{d_{\Delta}}\right)\tanh\left(\frac{L_{\text{w}}-x}{d_{\Delta}}\right),\label{eq:Delta_profile}
\end{eqnarray}
for $0<x<L_{\text{w}}$, i.e., a smooth profile that vanishes at the
ends of the nanowire. Here $\Delta_{0}$ is the value in the bulk
(i.e., right next or beneath the bulk SC), and $d_{\Delta}$ is an
adjustable parameter that controls the slope of the profile. As mentioned
above, a more rigorous treatment of this mean-field Hamiltonian should
involve a self-consistent determination of this profile, but for our
present purposes this simplification is well justified.%
{} In contrast to Ref. \cite{Takei13_Soft_gap}, here we only
consider the deterministic profile Eq. (\ref{eq:Delta_profile}) and
we neglect other random inhomogeneities in $\Delta\left(x\right)$
introduced by disorder. More details on disorder-induced SC pairing
potential fluctuations can be found in Ref. \cite{Takei13_Soft_gap}.

In the absence of disorder and in the uniform case (i.e., limit $\upsilon_{\mu}=d_{\Delta}=0$),
the Hamiltonian $H$ in the limit $L_{\text{w}}\rightarrow\infty$,
can be easily diagonalized in momentum space $k$. In that case, the
dispersion relation for the Bogoliubov quasiparticles is\cite{Sau10_Proposal_for_MF_in_semiconductor_heterojunction,Oreg'10}
$E_{k,\pm}^{2}=V_{x}^{2}+\Delta_{0}^{2}+\xi_{k}^{2}+\left(\alpha_{R}k\right)^{2}\pm2\sqrt{V_{x}^{2}\Delta_{0}^{2}+\xi_{k}^{2}\left[V_{x}^{2}+\left(\alpha_{R}k\right)^{2}\right]}$,
with $\xi_{k}=\hbar^{2}k^{2}/\left(2m\right)-\mu_{0}$. For given
values $\mu_{0},\Delta_{0}$ and $\alpha_{R}$, this model has a TQPT
as a function of magnetic field $V_{x}$ (i.e., the Zeeman spin splitting)
from a topologically trivial phase to a non-trivial phase with the
appearance of MBS localized at the ends of the nanowire at the critical
Zeeman field value $V_{x,c}=\sqrt{\Delta_{0}^{2}+\mu_{0}^{2}}$, as
originally shown by Sau \textit{et al.}\cite{Sau10_Proposal_for_MF_in_semiconductor_heterojunction}
In the presence of disorder and other spatial fluctuations of the
parameters in the model, the critical field $V_{x,c}$ typically shifts
to larger values and its value depends on the precise details of the
disorder realization.\cite{Brouwer11_Topological_SC_in_disorder_wires,DeGottardi11_MFs_with_disorder,Adagideli13_Topological_order_in_dirty_wires,Fregoso13_Electrical_detection_of_TQPT,DeGottardi_MFs_with_spatially_varying_potentials}
The determination of the critical field defining the TQPT is then
non-trivial and has to be done numerically for a given disorder realization.
This is the subject of the next section.

Finally, we mention that our NSN system is actually conceptually (and
perhaps practically too) simpler than the usual NSN system (where
the \textquoteleft{}S\textquoteright{} part is an intrinsic superconductor)
because of the proximate nature of the superconductivity induced in
the nanowire from the metallic superconductor underneath the semiconductor.
Thus various complications (e.g. dissipation, cooling, self-consistency,
nontrivial Fermi distribution, electron heating, etc.) which might
make the description of the usual NSN structures difficult\cite{Vercruyssen12_Non_equilibrium_NSN_wires}
are most likely irrelevant in our system, where the \textquoteleft{}S\textquoteright{}
part is the nanowire on a real superconductor, making our theoretical
description easier than that for the standard NSN structures with
\textquoteleft{}S\textquoteright{} being a real superconducting nanowire
connected to two normal metallic tunnel contacts.

\section{\label{sec:transfer}Thermal transport and topological phase diagram
of a dirty Majorana nanowire}

Let us now focus on the topological phase diagram of the disordered
Majorana nanowire. In order to make progress, we have discretized
the Hamiltonian in Eqs. (\ref{eq:H_0}) and (\ref{eq:H_SC}), and
obtained a $N-$site tight-binding model with the lattice parameter
$a$ (see Ref. \cite{Stanescu11_MFs_in_SM_nanowires})

\begin{eqnarray}
H_{\text{NW}} & = & -t\sum_{\langle lm\rangle,\sigma}c_{l,\sigma}^{\dagger}c_{m,\sigma}-\sum_{l,\sigma}c_{l,\sigma}^{\dagger}\left(\mu_{l}-V_{x}\hat{\boldsymbol{\sigma}}_{\sigma\sigma^{\prime}}^{x}\right)c_{l,\sigma^{\prime}}\nonumber \\
 & &+\sum_{l,\sigma}\left(i\alpha\ c_{l,\sigma}^{\dagger}\hat{\boldsymbol{\sigma}}_{\sigma\sigma^{\prime}}^{y}c_{l+1,\sigma^{\prime}}+\Delta_{l}c_{l\uparrow}^{\dagger}c_{l\downarrow}^{\dagger}+{\rm H.c.}\right),\label{eq:Hw}
\end{eqnarray}
where $c_{l,s}^{\dagger}$, $\mu_{l}$ and $\Delta_{l}$ are the discrete
versions of $\psi_{\sigma}^{\dagger}\left(x\right)$, $\mu\left(x\right)$
and $\Delta\left(x\right)$, respectively, and $t=\hbar^{2}/2m_{e}a^{2}$
is the effective hopping parameter. Here $\alpha=\sqrt{m\alpha_{R}^{2}/2}$
is the corresponding Rashba coupling parameter in the tight-binding
model. The first site at the left end corresponds to $l=1$ and the
final site at the right is $l=N$. 

We consider a \textit{single} distribution of $\mu_{l}$ (disorder
realization), and systematically vary its dispersion $\upsilon_{\mu}$
around the mean value $\mu_{0}$. As mentioned above, $\upsilon_{\mu}$
is not an experimentally tunable parameter, but it is useful and instructive
to visualize the topological phase diagram as a function of varying
disorder. Presumably, a fixed disorder realization is closer to the
experiment, where the semiconductor nanowire is in the mesoscopic
regime, and it is not clear that disorder necessarily self-averages
at the very low experimental temperatures. We mention that whether
the experimental temperatures are low enough so that the system is
not self-averaging (so that mesoscopic fluctuations are important
as one goes from one sample) is currently not known for the Majorana
experiments, and the issue of whether to ensemble average over disorder
realizations or not for quantitative comparison with experiments remains
open at this stage.

We compute the topological phase diagram of the \textit{isolated}
nanowire (i.e., in absence of the normal contacts) using the transfer-matrix
approach\cite{DeGottardi11_MFs_with_disorder,Fregoso13_Electrical_detection_of_TQPT}
for the model Hamiltonian Eq. (\ref{eq:Hw}), as a function of the
disorder strength $\upsilon_{\mu}$ and the external Zeeman field
$V_{x}$. Physically, the transfer matrix relates states in the left
end to states in the right end of the wire. This statement can be
made more precise introducing the Majorana basis $c_{l,\sigma}=\left(a_{\sigma,l}+ib_{\sigma,l}\right)/\sqrt{2}$,
where the Majorana operators obey the anti-commutation relations $\left\{ a_{\sigma,l},a_{s,m}\right\} =\left\{ b_{\sigma,l},b_{s,m}\right\} =\delta_{l,m}\delta_{\sigma,s}$
and zero otherwise. In terms of these operators, a generic eigenmode
$\Psi$ of $H_{\text{NW}}$ satisfying the eigenvalue equation $H_{\text{NW}}\Psi=E\Psi$
can be written as

\begin{eqnarray}
\Psi & =&\sum_{l=1}^{N}\left(\gamma_{\uparrow,l}a_{\uparrow,l}+\gamma_{\downarrow,l}a_{\downarrow,l}+\eta_{\uparrow,l}b_{\uparrow,l}+\eta_{\downarrow,l}b_{\downarrow,l}\right),\label{eq:psi}
\end{eqnarray}
with real coefficients $\gamma_{\sigma,l}$ and $\eta_{\sigma,l}$.
At $E=0$, defining the matrices $\boldsymbol{\kappa}=\left(\begin{array}{cc}
t & -\alpha\\
\alpha & t
\end{array}\right)$, $\mathbf{u}_{l}=\left(\begin{array}{cc}
\mu_{l} & \Delta_{l}-V_{x}\\
-\Delta_{l}-V_{x} & \mu_{l}
\end{array}\right)$ and the vector of coefficients $\vec{\psi}_{l}=\left(\gamma_{\uparrow,l},\gamma_{\downarrow,l}\right)^{T}$,
the above eigenvalue equation can be written as $0=\boldsymbol{\kappa}^{\dagger}\vec{\psi}_{l-1}+\boldsymbol{\kappa}\vec{\psi}_{l+1}+\mathbf{u}_{l}\vec{\psi}_{l}$,
and from here we obtain the transfer equation

\begin{eqnarray}
\left(\begin{array}{c}
\vec{\psi}_{l+1}\\
\boldsymbol{\kappa}^{\dagger}\vec{\psi}_{l}
\end{array}\right) & =\mathbf{M}_{l}\left(\begin{array}{c}
\vec{\psi}_{l}\\
\boldsymbol{\kappa}^{\dagger}\vec{\psi}_{l-1}
\end{array}\right),\label{eq:transfer_equation}
\end{eqnarray}
where 

\begin{eqnarray}
\mathbf{M}_{l} & \equiv\left(\begin{array}{cc}
-\boldsymbol{\kappa}^{-1}\mathbf{u}_{l} & -\boldsymbol{\kappa}^{-1}\\
\boldsymbol{\kappa}^{\dagger} & 0
\end{array}\right),\nonumber \\
 & =\left(\begin{array}{cccc}
\frac{-t\mu_{l}+\alpha\left(V_{x}-\Delta_{l}\right)}{t^{2}+\alpha^{2}} & \frac{-\alpha\mu_{l}+t\left(V_{x}+\Delta_{l}\right)}{t^{2}+\alpha^{2}} & \frac{t}{t^{2}+\alpha^{2}} & \frac{\alpha}{t^{2}+\alpha^{2}}\\
\frac{\alpha\mu_{l}+t\left(V_{x}-\Delta_{l}\right)}{t^{2}+\alpha^{2}} & \frac{-t\mu_{l}-\alpha\left(V_{x}+\Delta_{l}\right)}{t^{2}+\alpha^{2}} & \frac{-\alpha}{t^{2}+\alpha^{2}} & \frac{t}{t^{2}+\alpha^{2}}\\
-t & -\alpha & 0 & 0\\
\alpha & -t & 0 & 0
\end{array}\right)\label{eq:Ml}
\end{eqnarray}
is the $l-$th transfer matrix relating the vectors $\vec{\psi}_{l+1}$
and $\vec{\psi}_{l-1}$. Then, the full transfer matrix of the nanowire,
from site $l=1$ to site $l=N$, is simply given by $\mathbf{M}=\prod_{l=1}^{N}\mathbf{M}_{l}$.
The eigenvalues of $\mathbf{M}$ can be written as $e^{\pm N\lambda_{n}}$,
where $\lambda_{n}$ are the (dimensionless) ``Lyapunov exponents''
of the system,\cite{Beenakker1997} which represent the inverse of
the localization length. The connection to localization properties
are better understood recalling that the transmission probability
from site $1$ to site $N$ is $\mathcal{T}_{1N}=\sum_{n=1}^{4}\mathcal{T}_{n},$
with

\begin{eqnarray}
\mathcal{T}_{n} & =&\cosh^{-2}\left(N\lambda_{n}\right),\label{eq:Tn}
\end{eqnarray}
the transmission eigenvalue corresponding to the $n-$th channel .\cite{Beenakker1997} 

The connection between the localization and the topological properties
of a ``dirty'' class D nanowire was made explicit by Akhmerov \textit{et
al} ,\cite{Akhmerov11_Quantized_conductance_in_disordered_wire} who
obtained the topological invariant $Q=\text{sign}(\prod_{n=1}^{2M}\tanh\lambda_{n})$, with $M$ the number of channels in the the wire.
These authors have shown that in the clean case this topological invariant
actually reduces to the one derived by Kitaev which is given in terms
of the Pfaffian of the Hamiltonian in momentum space.\cite{kitaev2001}
Here we see explicitly that $Q$ changes sign when one of the Lyapunov
exponents vanishes and changes sign. This signals the TQPT. As discussed
in Refs. \cite{Motrunich01_Disorder_in_topological_1D_SC, Akhmerov11_Quantized_conductance_in_disordered_wire},
the TQPT of a class D SC corresponds to a \textit{delocalization point}
for zero-energy particles, i.e., one of the Lyapunov exponents $\lambda_{n}$
vanishes and changes sign at the TQPT inducing a ``perfect'' transmission
probability $\mathcal{T}_{n}=1$. Everywhere else in the parameter
space the system is localized at zero energy, i.e., all $\lambda_{n}$
are finite. This crucial result will be addressed in detail in Section
\ref{sec:intuitive}. For the moment, we can check that this idea
also works in the clean case: for a clean nanowire, sufficiently close
to the TQPT on the topological side, the MBS wavefunctions are localized
within the SC correlation length $\xi_{\text{clean}}\simeq\hbar v_{F}/\Delta\left(V_{x}\right)$,
where $\Delta\left(V_{x}\right)$ is the effective SC quasiparticle
gap controlled by the Zeeman field. The TQPT is reached at the critical
field $V_{x,c}=\sqrt{\Delta_{0}^{2}+\mu_{0}^{2}}$, where the quasiparticle
gap $\Delta\left(V_{x,c}\right)\rightarrow0$ and the localization
length $\xi_{\text{clean}}\rightarrow\infty$. When $\xi_{\text{clean}}\simeq L_{\text{w}}$,
the MBS localized at opposite ends can ``see'' each other and overlap
forming a Majorana ``channel'' that connects the left and the right
end. Therefore, for a clean system near the TQPT the smallest Lyapunov
exponent is $\lambda_{\text{clean}}\propto\xi_{\text{clean}}^{-1}$.
Since the Majorana channel has equal contributions of electrons and
holes at $E=0$, the current sustained by electron-like states exactly
cancels the current of hole-like states, and the total electric current
vanish. Therefore, the perfectly quantized transmission coefficient
$\mathcal{T}_{n}=1$ occurring at the TQPT is physically related to
the thermal conductance (and not to the electrical conductance). We
will return to this point in Sec. \ref{sec:transport}.

In Fig. \ref{fig:GLL_GRR_pdiag}(c) we show a 2D color map of the
thermal transmission coefficient $\mathcal{T}_{1N}$ for a dirty wire
as a function of disorder ``strength'' $\upsilon_{\mu}$ and applied
Zeeman field $V_{x}$, fixing all other parameters (chemical potential,
pair-potential profile, etc.) according to Table \ref{tab:parametersTB}.
These parameters correspond exactly to those used in Figs. \ref{fig:GLL_GRR_pdiag}(a)
and \ref{fig:GLL_GRR_pdiag}(b) for the\textit{ same configuration
of disorder potential}, and each dot corresponds to each one of the
curves in those figures. The blue regions indicate the points for
which the transmission coefficient is close to the maximal value $\mathcal{T}_{1N}=1$,
and therefore indicate the approximate location of the TQPT. Therefore,
Fig. \ref{fig:GLL_GRR_pdiag}(c) allows to determine the phase boundary
separating the topological from the non-topological region. Note that
this boundary has an intrinsic width which scales as $\propto1/N\propto1/L_{\text{w}}$.
More precisely, as we will see in Sec. \ref{sec:intuitive}, the width
corresponds to the Thouless energy $\hbar v_{F}/L_{\text{w}}$. For
the parameters in Table \ref{tab:parametersTB} and in the absence
of disorder, we estimate an upper bound $L_{\text{w}}/\xi\simeq15$,
where we have used the estimation for the minimal value of the SC
correlation length $\xi_{\text{clean}}=\hbar v_{F}/\Delta_{0}\simeq20\ a$
and $L_{\text{w}}=300\ a$.

Contrasting Figs. \ref{fig:GLL_GRR_pdiag}(a), \ref{fig:GLL_GRR_pdiag}(b)
and \ref{fig:GLL_GRR_pdiag}(c), we note that the four curves
on the top correspond to dots in (c) which are closer to the topological phase
boundary, where the topological protection is expected to be more
fragile. This seems to be in agreement with the fact that Fig. \ref{fig:GLL_GRR_pdiag}(a)
shows a splitting in the ZBP. The ZBP appearing in the corresponding
curves in Figs. \ref{fig:GLL_GRR_pdiag}(b) and the apparent inconsistency
with Fig. \ref{fig:GLL_GRR_pdiag}(a) (i.e., peaks not correlated)
will be addressed and discussed in the next section. In contrast,
the four central curves in both figure (a) and (b) show a more robust
ZBP, which is consistent with the corresponding points in Fig. \ref{fig:GLL_GRR_pdiag}(c)
located further from the boundary. As we see, this analysis showing
all curves ``side-to-side'' is potentially helpful to interpret
the experimental transport results. A natural question arises: is
it possible to access the information in (c) \textit{experimentally}?
This will be the subject of the next sections. 

\begin{table*}[t]
\begin{tabular}{ccc}
\hline 
Parameter  & Value in InSb (if applicable) & TB equivalent\tabularnewline
\hline 
Wire length & $L_{\text{w}}=2\ \mu\text{m}$ & $N=300$\tabularnewline
Mass & $m=0.015\ m_{e}$ & $m=\left(2ta^{2}\right)^{-1}$\tabularnewline
Chemical Potential & not known & $\mu_{0}=-1.72\ t$\tabularnewline
Bulk pairing potential & $\Delta_{0}=250\ \mu$eV & $\Delta_{0}=0.05\ t$\tabularnewline
Rashba spin-orbit coupling & $\alpha_{R}=0.2\ \text{eV}.\AA$ & $\alpha=\sqrt{m\alpha_{R}^{2}t/2}=0.15\ t$\tabularnewline
Slope of pairing profile & not applicable & $d_{\Delta}=30\ a$\tabularnewline
\hline 
\end{tabular}

\caption{\label{tab:parametersTB}Parameters used in the model (\ref{eq:Hw})
in the numerical simulations in Figs. \ref{fig:GLL_GRR_pdiag}, \ref{fig:Gall}
and \ref{fig:dGLL}. The hopping parameter $t=1$ meV has been chosen
to reproduce a ratio $L_{\text{w}}/\xi_{\text{clean}}\approx15$.
The average chemical potential $\mu_{0}$ has been chosen to reproduce
the reported experimental value of the critical Zeeman field\cite{Mourik12_Signatures_of_MF}
(i..e, $B\simeq250$ mT) using the formula\cite{Sau10_Proposal_for_MF_in_semiconductor_heterojunction}
$V_{x,c}=\sqrt{\Delta_{0}^{2}+\mu_{0}^{2}}$ and assuming weak disorder.}
 
\end{table*}

\section{\label{sec:transport}Electronic transport properties in the NSN
configuration}

\begin{figure}[t]
\begin{centering}
\includegraphics[bb=20bp 150bp 650bp 550bp,clip,scale=0.37]{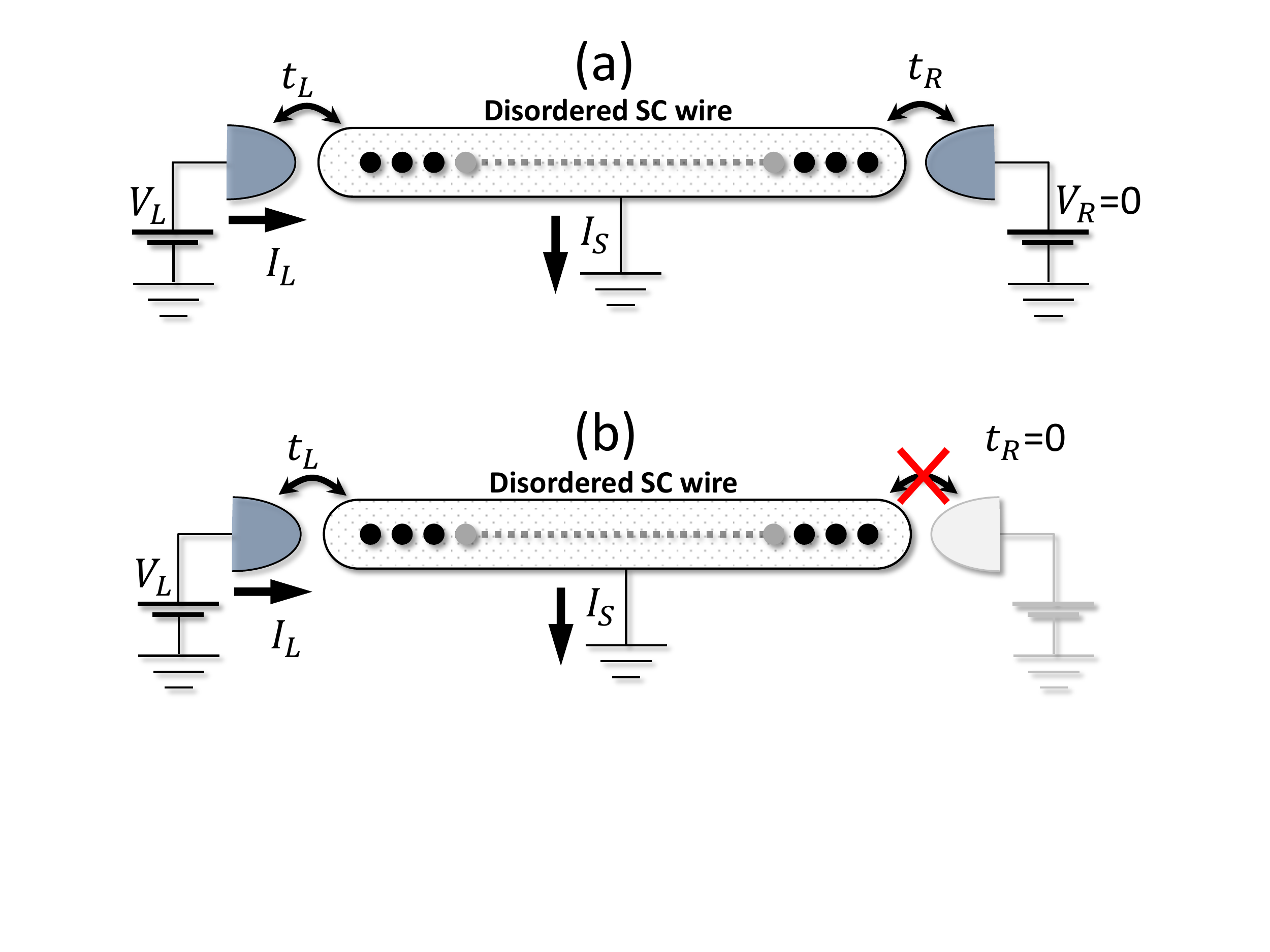}\caption{(a) Schematic view of an NSN circuit. (b) Assuming that the coupling
to the leads can be controlled \textit{in situ} experimentally {[}e.g.,
using pinch off gates (not shown here){]}, the system can be effectively
disconnected from the right lead and turned effectively into an NS
junction. \label{fig:NSN_circuit-1}}
\end{centering}
\end{figure}
We now turn to quantities with more relevance to current experimental
measurements. To that end, we introduce a term in the Hamiltonian
describing the coupling to external normal leads {[}see Fig.~\ref{fig:NSN_circuit-1}(a){]}

\begin{eqnarray}
H_{\text{mix}} & =&\sum_{\sigma}\left(t_{L}d_{Lk,\sigma}^{\dagger}c_{1,\sigma}+t_{R}d_{Rk,\sigma}^{\dagger}c_{N,\sigma}\right)+\text{H.c.,}\label{eq:H_mix}
\end{eqnarray}
where the term where $t_{L\left(R\right)}$ is the coupling to the
left (right) lead and $d_{L\left(R\right)k,s}^{\dagger}$ is the corresponding
creation operator for fermions with quantum number $k$ and spin $\sigma$.
The external leads are modeled as large Fermi liquids with Hamiltonian
$H_{\text{lead},j}=\sum_{k,\sigma}\epsilon_{k}d_{j,k,\sigma}^{\dagger}d_{j,k,\sigma}$,
where $j=\left\{ L,R\right\} $. We assume that each lead is in equilibrium
at a chemical potential $\mu_{j}=eV_{j}$ controlled by external voltages,
and that the SC nanowire is grounded. The expression for the electric
current flowing through the contacts is $I_{j}=e\langle dN_{j}/dt\rangle=ie\langle[H,N_{j}]\rangle/\hbar=ie\langle[H_{\text{mix}},N_{j}]\rangle/\hbar$,
which can be written using equations of motion in terms of the Green's
function in the nanowire \cite{Cuevas96_Hamiltonian_approach_to_SC_contacts,meir92}.
The excess current $I_{S}$ flowing to ground through the bulk of
the SC wire ensures the average conservation of charge $I_{L}+I_{R}+I_{S}=0$.
The conductance matrix of the NSN system Eq. (\ref{eq:Gmatrix}) can
be expressed as


\begin{eqnarray}
G_{LL} & =&\frac{e^{2}}{h}\int_{-\infty}^{\infty}d\omega\;\left[-\frac{dn_{L}\left(\omega\right)}{d\left(eV_{L}\right)}\right]\text{Tr }\left[2\mathbf{r}_{eh}^{LL}\left(\mathbf{r}_{eh}^{LL}\right)^{\dagger}+\mathbf{t}_{ee}^{LR}\left(\mathbf{t}_{ee}^{LR}\right)^{\dagger}+\mathbf{t}_{eh}^{LR}\left(\mathbf{t}_{eh}^{LR}\right)^{\dagger}\right]_{\omega},\label{eq:GLL_final_symm-1}\\
G_{LR} & =&\frac{e^{2}}{h}\int_{-\infty}^{\infty}d\omega\;\left[\frac{dn_{R}\left(\omega\right)}{d\left(eV_{R}\right)}\right]\text{Tr }\left[\mathbf{t}_{ee}^{LR}\left(\mathbf{t}_{ee}^{LR}\right)^{\dagger}-\mathbf{t}_{eh}^{LR}\left(\mathbf{t}_{eh}^{LR}\right)^{\dagger}\right]_{\omega},\label{eq:GLR_final_symm-1}\\
G_{RL} & =&\frac{e^{2}}{h}\int_{-\infty}^{\infty}d\omega\;\left[\frac{dn_{L}\left(\omega\right)}{d\left(eV_{L}\right)}\right]\text{Tr }\left[\mathbf{t}_{ee}^{RL}\left(\mathbf{t}_{ee}^{RL}\right)^{\dagger}-\mathbf{t}_{eh}^{RL}\left(\mathbf{t}_{eh}^{RL}\right)^{\dagger}\right]_{\omega},\label{eq:GRL_final_symm-1}\\
G_{RR} & =&\frac{e^{2}}{h}\int_{-\infty}^{\infty}d\omega\;\left[-\frac{dn_{R}\left(\omega\right)}{d\left(eV_{R}\right)}\right]\text{Tr }\left[2\mathbf{r}_{eh}^{RR}\left(\mathbf{r}_{eh}^{RR}\right)^{\dagger}+\mathbf{t}_{ee}^{RL}\left(\mathbf{t}_{ee}^{RL}\right)^{\dagger}+\mathbf{t}_{eh}^{RL}\left(\mathbf{t}_{eh}^{RL}\right)^{\dagger}\right]_{\omega},\label{eq:GRR_final_symm-1}
\end{eqnarray}

These formulas are standard (see Refs. \cite{Blonder1982_BTK_paper,Anantram96_Andreev_scattering,Prada12_Transport_through_NS_junctions_with_MBS,Fregoso13_Electrical_detection_of_TQPT})
and we do not derive them here. The reader will find more details
in the aforementioned references and in Appendix \ref{sec:appendix}.
We have defined the normal reflection and transmission matrices (i.e.,
with subindex ``$ee$'' or ``electron-electron'')

\begin{eqnarray*}
\left[\mathbf{r}_{ee}^{LL}\left(\omega\right)\right]_{\sigma,\sigma^{\prime}} & =&\gamma_{L}\left(\omega\right)g_{1\sigma,1\sigma^{\prime}}^{r}\left(\omega\right),\\
\left[\mathbf{r}_{ee}^{RR}\left(\omega\right)\right]_{\sigma,\sigma^{\prime}} & =&\gamma_{R}\left(\omega\right)g_{N\sigma,N\sigma^{\prime}}^{r}\left(\omega\right),\\
\left[\mathbf{t}_{ee}^{LR}\left(\omega\right)\right]_{\sigma,\sigma^{\prime}} & =&\sqrt{\gamma_{L}\left(\omega\right)\gamma_{R}\left(\omega\right)}g_{1\sigma,N\sigma^{\prime}}^{r}\left(\omega\right),
\end{eqnarray*}
and the Andreev reflection and transmission matrices (i.e., with subindex
``$eh$'' or ``electron-hole'')
\begin{eqnarray*}
\left[\mathbf{r}_{eh}^{LL}\left(\omega\right)\right]_{\sigma,\sigma^{\prime}} & =&\gamma_{L}\left(\omega\right)f_{1\sigma,1\sigma^{\prime}}^{r}\left(\omega\right),\\
\left[\mathbf{r}_{eh}^{RR}\left(\omega\right)\right]_{\sigma,\sigma^{\prime}} & =&\gamma_{R}\left(\omega\right)f_{N\sigma,N\sigma^{\prime}}^{r}\left(\omega\right),\\
\left[\mathbf{t}_{eh}^{LR}\left(\omega\right)\right]_{\sigma,\sigma^{\prime}} & =&\sqrt{\gamma_{L}\left(\omega\right)\gamma_{R}\left(\omega\right)}f_{1\sigma,N\sigma^{\prime}}^{r}\left(\omega\right),
\end{eqnarray*}
where $g_{ls,ms^{\prime}}^{r}\left(\omega\right)$ and $f_{ls,ms^{\prime}}^{r}\left(\omega\right)$
are the normal and anomalous retarded Green's functions\cite{fetter}
in the nanowire respectively (see Appendix \ref{sec:appendix}). We
have also defined the effective couplings to the leads 
\begin{eqnarray}
\gamma_{j}\left(\omega\right) & =&2\pi t_{j}^{2}\rho_{j}^{0}\left(\omega\right)\ \ \ \left(j=L,R\right),\label{eq:gamma_j}
\end{eqnarray}
where $\rho_{j}^{0}\left(\omega\right)$ is the density of states
in the $j-$lead. Assuming a large bandwidth in the normal contacts,
in the following we set $\rho_{j}^{0}\left(\omega\right)=\rho_{j}^{0}\left(0\right)$,
the value at the Fermi level. 

Let us analyze the physical content of Eqs. (\ref{eq:GLL_final_symm-1})-(\ref{eq:GRR_final_symm-1}).
We first focus on the ``local'' conductances Eqs. (\ref{eq:GLL_final_symm-1})
and (\ref{eq:GRR_final_symm-1}). In these expressions, the first
term corresponds to the local contribution $2\text{Tr }\left[\mathbf{r}_{eh}^{LL}\left(\mathbf{r}_{eh}^{LL}\right)^{\dagger}\right]_{\omega}$
and $2\text{Tr }\left[\mathbf{r}_{eh}^{RR}\left(\mathbf{r}_{eh}^{RR}\right)^{\dagger}\right]_{\omega}$,
i.e., the Andreev reflection probability at the left and right lead,
respectively. These terms are the only terms appearing in the case
of NS or SN contacts,\cite{Law09} and they already contain the information
about the presence of a MBS localized at the corresponding end. From
this perspective, the quantized value of the conductance $2e^{2}/h$
corresponds to a ``perfect'' Andreev reflection $\text{Tr }\left[\mathbf{r}_{eh}^{LL}\left(\mathbf{r}_{eh}^{LL}\right)^{\dagger}\right]_{\omega=0}=1$
at $T=0$, due to the presence of the MBS. However, note that in Eqs.
(\ref{eq:GLL_final_symm-1}) and (\ref{eq:GRR_final_symm-1}) we also
encounter a \textit{non-local} contribution $\text{Tr }\left[\mathbf{t}_{ee}^{LR}\left(\mathbf{t}_{ee}^{LR}\right)^{\dagger}+\mathbf{t}_{eh}^{LR}\left(\mathbf{t}_{eh}^{LR}\right)^{\dagger}\right]_{\omega}$,
which physically corresponds to particles that travel from one to
the other end of the wire, and return to the original lead with information
about the opposite lead. These non-local terms are present only in
the NSN junctions, and not in the simple NS configuration so far studied
extensively in the literature. Our primary motivation for considering
NSN junctions (with the 'S' part being the nanowire carrying MBS under
suitable conditions) is to study the effect of these non-local terms
in the transport experiments, since non locality is the key concept
underlying MBS in the topological phase. Therefore, this contribution
must be proportional to the electron-electron and electron-hole \textit{transmission}
coefficients, $\text{Tr }\left[\mathbf{t}_{ee}^{LR}\left(\mathbf{t}_{ee}^{LR}\right)^{\dagger}\right]_{\omega}$
and $\text{Tr }\left[\mathbf{t}_{eh}^{LR}\left(\mathbf{t}_{eh}^{LR}\right)^{\dagger}\right]_{\omega}$
respectively, and vanishes if either $\gamma_{L}$ or $\gamma_{R}$
vanishes by, e.g., ``pinching off'' one of the quantum point contacts
using underlying gates (i.e., pinch off gates) {[}see Fig. \ref{fig:NSN_circuit-1}(b){]}.
Note that the presence of such a non-local contribution is expected
in multi-terminal phase-coherent mesoscopic systems.\cite{Buttiker86_Landauer_Buttiker_paper} 

\begin{figure*}[t]
\begin{centering}
\includegraphics[bb=0bp 150bp 650bp 650bp,clip,scale=0.6]{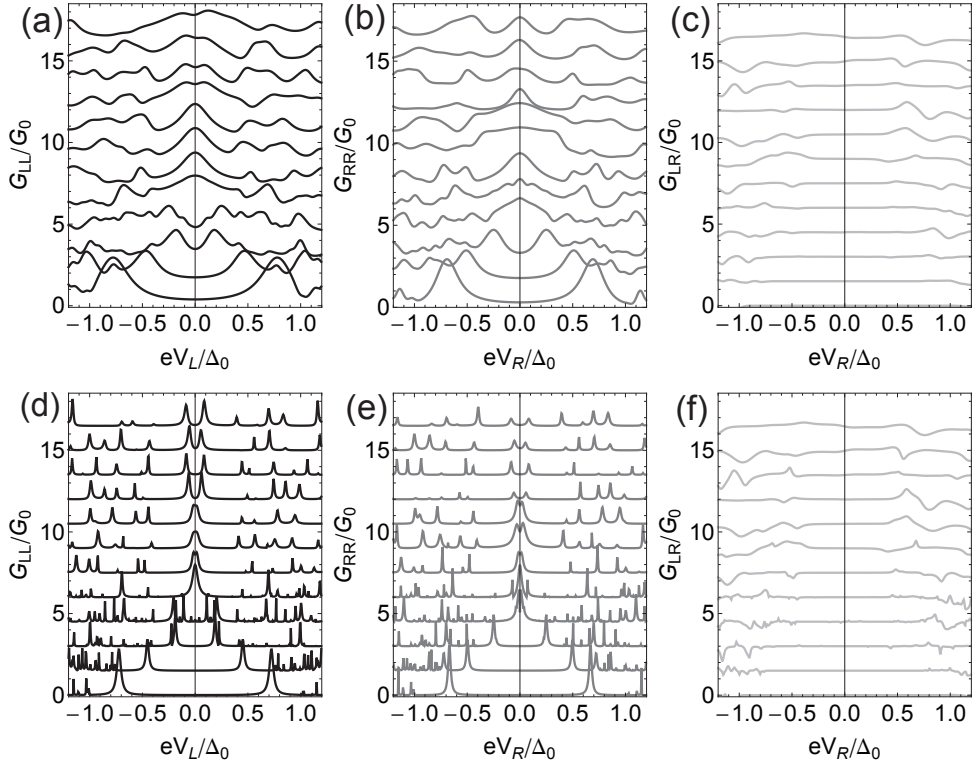}\caption{\label{fig:Gall}(a) Differential conductances $G_{LL}$ vs $V_{L}$
, (b) $G_{RR}$ vs $V_{R}$ and (c) $G_{LR}$ vs $V_{R}$. All the
curves in the top {[}i.e, (a), (b) and (c){]} have been computed at
a temperature $T/\Delta_{0}=0.02$ (corresponding to the experimental
temperature $T_{\text{exp}}\approx30$ mK) and for effective couplings
to the leads $\gamma_{L}=0.85\ t$ and $\gamma_{R}=0.95\ t$. Figs.
\ref{fig:Gall}(a) and (b) correspond to Figs. \ref{fig:GLL_GRR_pdiag}(a)
and (b) and exemplify the potential difficulties in detecting the
TQPT in a disordered Majorana wire, with non-topological ZBPs appears
at finite Zeeman field. In the bottom plots {[}i.e, (d), (e) and (f){]}
we show the same quantities computed at much lower a temperature $T/\Delta_{0}=2\times10^{-4}$
(below experimental capabilities) and for effective couplings to the
leads $\gamma_{L}=\gamma_{R}=0.1\ t$. In these conditions, the widths
of the conductance peaks decrease dramatically, revealing the splitting
of the ZBPs due to disorder.}
\end{centering}
\end{figure*}
Interestingly, the thermal conductance $G_{\text{th}}$ across the
wire\cite{landauer57,Buttiker85_Many_channel_conductance,Buttiker86_Landauer_Buttiker_paper} 

\begin{eqnarray}
G_{\text{th}} & =&G_{\text{th},0}\text{Tr }\left[\mathbf{t}_{ee}^{LR}\left(\mathbf{t}_{ee}^{LR}\right)^{\dagger}+\mathbf{t}_{eh}^{LR}\left(\mathbf{t}_{eh}^{LR}\right)^{\dagger}\right]_{\omega},\label{eq:G_th}
\end{eqnarray}
where $G_{\text{th},0}=\pi^{2}k_{B}^{2}T/6h$ is the thermal quantum
of conductance, is closely connected to the non-local contribution
in Eqs. (\ref{eq:GLL_final_symm-1}) and (\ref{eq:GRR_final_symm-1}),
and allows to make a link with our previous discussion in Sec. \ref{sec:transfer}.
The connection with the thermal transmission probability $\mathcal{T}_{1N}$
at zero energy is $\mathcal{T}_{1N}=2\text{Tr }\left[\mathbf{t}_{ee}^{LR}\left(\mathbf{t}_{ee}^{LR}\right)^{\dagger}+\mathbf{t}_{eh}^{LR}\left(\mathbf{t}_{eh}^{LR}\right)^{\dagger}\right]_{\omega=0}$.
In principle, the information about the TQPT is contained in this
expression. However, such thermal measurements are in general very
challenging experimentally, and we need to come up with a different
approach which is experimentally feasible. In particular, it is desirable
to use electrical measurements (i.e., electrical conductance) for
observing the non-local MBS correlations at the TQPT and beyond. 

We now briefly discuss Eqs. (\ref{eq:GLR_final_symm-1}) and (\ref{eq:GRL_final_symm-1})
(i.e., the so-called \textit{transconductances} $G_{LR}$ and $G_{RL}$, which obey $G_{RL}=G_{LR}$),
where a more explicit difference with respect to the NS geometry appears.
Physically, the minus sign in these expressions appears because while
electrons contribute with a plus sign to the transport, a hole will
contribute a minus sign. As discussed previously in Sec. \ref{sec:transfer},
note that if the system is in the topological phase with end-MBS,
particle-hole symmetry dictates that the contributions $\text{Tr }\left[\mathbf{t}_{ee}^{LR}\left(\mathbf{t}_{ee}^{LR}\right)^{\dagger}\right]_{\omega=0}$
and $\text{Tr }\left[\mathbf{t}_{eh}^{LR}\left(\mathbf{t}_{eh}^{LR}\right)^{\dagger}\right]_{\omega=0}$
must be identical, and therefore the transconductance must vanish.\cite{Akhmerov11_Quantized_conductance_in_disordered_wire}
This might seem to rule out the possibility to see the TQPT via electrical
measurements of $G_{LR}$. %

In Figs. \ref{fig:Gall}(a) and \ref{fig:Gall}(b) we reproduce the
same Figs. \ref{fig:GLL_GRR_pdiag}(a) and \ref{fig:GLL_GRR_pdiag}(b),
computed for the parameters in Table \ref{tab:parametersTB}, at a
temperature $T/\Delta_{0}=0.02$ (which approximately corresponds
to the experimental temperature $T_{\text{exp}}\approx60$ mK in Ref.
\cite{Mourik12_Signatures_of_MF}), and where we have assumed
effective couplings $\gamma_{L}=0.85\ t$ and $\gamma_{R}=0.95\ t$,
corresponding to an open wire condition (i.e., ``good'' electrical
contact with the leads). In Fig. \ref{fig:Gall}(c) we present a plot
for $G_{LR}$ vs $V_{R}$ for the same parameters. Note that this
quantity is rather featureless, and is vanishingly small near zero
bias as expected. %
{} For comparison, in Figs. \ref{fig:Gall}(d), \ref{fig:Gall}(e) and
\ref{fig:Gall}(f) we show $G_{LL}$, $G_{RR}$ and $G_{LR}$, respectively,
for the same parameters but for a much lower temperature $T/\Delta_{0}=2\times10^{-4}$
and smaller couplings $\gamma_{L}=\gamma_{R}=0.1\ t$ . In these conditions
the thermal and quasiparticle broadenings decrease dramatically and
we realize that the preliminary information about the ZBPs in Figs.
\ref{fig:Gall}(a) and \ref{fig:Gall}(b) is misleading: the system
does not have zero-bias excitations and the peaks are actually split
(rather than being a single zero energy peak) in Figs. \ref{fig:Gall}(d)
and \ref{fig:Gall}(e) at low temperatures and at low transparency
of the contacts. This picture is actually consistent with Fig. \ref{fig:GLL_GRR_pdiag}(c),
where the dots corresponding to the largest magnetic fields are very
close to the topological phase boundary, and therefore we expect the
MBS to recombine into Dirac fermions and consequently the peaks to
shift away from zero bias voltage. This allows to interpret the uncorrelated
ZBPs for $G_{LL}$ and $G_{RR}$. The results in Fig. \ref{fig:Gall}
shows that already for the simple model of Eq. (\ref{eq:Hw}), detection
of a ``true'' Majorana ZBP based only on the information about the
local conductances might be very tricky.\cite{Liu12_ZBP_in_Majorana_wires_with_and_without_MZBSs}
Therefore, the presence of ZBPs in the local conductances $G_{LL}$
and $G_{RR}$ cannot by itself be considered as a ``smoking-gun''
evidence of the Majorana scenario without some critical considerations
of the correlations in the existence of these ZBPs arising from the
two end conductances.

\section{\label{sec:pdiag}Electrical detection of  topological phase transitions in the NSN configuration}

As mentioned before, in the case of clean wires, the TQPT should be
observed in the closing and reopening of the gap of electronic excitations
in the nanowire. This re-organization of the fermionic spectrum is
necessary in order to accommodate a new MBS at zero energy. However,
the experiments so far have been unable to report any definitive closing
of the gap. It has been suggested that this negative results might
originate because while the tunneling occurs at the end of the nanowire,
the information about the gap-closing is contained in wavefunctions
with most of the weight in the bulk of the nanowire. Therefore,
measurements of the LDOS in the middle of the wire\cite{Stanescu12_To_close_or_not_to_close}, capacitive measurements of the total DOS\cite{Appelbaum13_Gap_closing_in_Majorana_NWs}, or phase-locked magnetoconductance oscillations in flux-biased topological Josephson junctions \cite{Diez13_Magnetoconductance_oscillations_as_a_probe_of_MFs}
should reveal this gap-closing occurring at the TQPT, but no experimental
evidence of these predictions have been reported so far in nanowires contacted at the ends. 

In addition to characterizing the transport properties of disordered
NSN Majorana wires, another goal of the present work is to explore
experimental proposals to determine the topological phase diagram.
We believe that the NSN geometry offers an interesting possibility
to achieve this goal, and to provide information about the topological
stability of the MBS. In Sec. \ref{sec:transfer} we stressed that
the TQPT in Majorana wires corresponds to a delocalization point at
zero energy, a fact that can be detected in the thermal transmission
probability across the system. On the other hand, in Sec. \ref{sec:transport}
we showed that in the NSN geometry, the local conductance of a phase-coherent
Majorana nanowire depends on the non-local transmission probability
$\text{Tr }\left[\mathbf{t}_{ee}^{LR}\left(\mathbf{t}_{ee}^{LR}\right)^{\dagger}+\mathbf{t}_{eh}^{LR}\left(\mathbf{t}_{eh}^{LR}\right)^{\dagger}\right]_{\omega}$,
in addition to the local Andreev reflection coefficient, which exactly
corresponds to the (dimensionless) thermal transport at energy $\omega$
{[}see Eq. (\ref{eq:G_th}){]}. This enables mapping out the topological
phase diagram by purely \textit{electrical} measurements.\cite{Fregoso13_Electrical_detection_of_TQPT}
In this section we provide more details in the way the non-local information
could be extracted in the NSN configuration in order to obtain the
topological phase diagram.

Following Ref. \cite{Fregoso13_Electrical_detection_of_TQPT},
we define the following quantity

\begin{eqnarray}
\Delta G_{jj}\left(0\right) & \equiv & G_{jj}\left(0\right)-G_{jj}^{\prime}\left(0\right),\label{eq:dGjj}
\end{eqnarray}
i.e., the difference of local zero-bias conductances computed for
different values of couplings to the \textit{opposite} lead, while
keeping all other parameters fixed. The zero-bias conductance at one
end $G_{jj}\left(0\right)$ is computed for a given value of $\gamma_{\bar{j}}$
(with compact notation $\bar{L}=R$ and $\bar{R}=L$) and $G_{jj}^{\prime}\left(0\right)$
is computed for a different value $\gamma_{\bar{j}}^{\prime}$. From
the experimental point of view, this means using $\gamma_{R}$ and
$\gamma_{L}$ as tuning parameters, something that could be achieved
varying the pinch-off gates underneath the ends of the nanowire.\cite{Mourik12_Signatures_of_MF,Das12_Evidence_of_MFs,Deng12_ZBP_in_Majorana_NW,Rokhinson2012,Finck13_ZBP_in_hybrid_NW_SC_device,Churchill2013}
This constitutes a new experimental knob which has not been explored
so far in the Majorana experiment. Note that this quantity {[}as defined
by Eq. (\ref{eq:dGjj}){]}, being a difference, is not quantized and
can take either positive or negative values. For this reason in what
follows we will take the absolute value.

A priori, it might seem counter-intuitive that the transport through
a disordered medium could be influenced by the change of a boundary
condition at the far-end. However, this intuition is typically built
upon the more usual case of trivial Anderson-localized 1D systems,
where any amount of disorder localizes the wavefunctions and therefore
any object placed at distances larger than the localization length
$\xi_{\text{loc}}$ has essentially no effect. The crucial difference
with class D conductors is that $\xi_{\text{loc}}\propto\lambda_{n}^{-1}\rightarrow\infty$
at the TQPT (since the gap closes here), i.e., the delocalization
point. Therefore, assuming that $L_{\text{w}}<L_{\phi}$, where $L_{\phi}$
is the phase-relaxation length, the aforementioned intuition is usually
correct \textit{except} at the TQPT. The physical idea behind using
$\Delta G_{jj}\left(0\right)$ as an indicator of the TQPT can be
seen quite simply in the extreme case when $\gamma_{\bar{j}}^{\prime}=0$.
In this case, in Eqs. (\ref{eq:GLL_final_symm-1}) and (\ref{eq:GRR_final_symm-1})
for $G_{LL}^{\prime}$ and $G_{RR}^{\prime}$ respectively, one completely
suppresses the coupling to the opposite lead and the transmission
coefficients vanish. The remaining part (i.e., Andreev reflection
$2\text{Tr }\left[\mathbf{r}_{eh}^{jj}\left(\mathbf{r}_{eh}^{jj}\right)^{\dagger}\right]_{\omega}$)
is a purely local contribution. Therefore, Eq. (\ref{eq:dGjj}) must
correspond to a non-local contribution which contains information
about the TQPT. This statement is not entirely correct because modifying
the coupling $\gamma_{\bar{j}}$ to $\gamma_{\bar{j}}^{\prime}$ also
modifies the local Andreev reflection coefficient through the local
anomalous Green's functions $f_{1s,1s^{\prime}}^{r}\left(\omega\right)$,
which contains information about the entire system. Only in the perturbative
limit where $\delta\gamma_{\bar{j}}\equiv\gamma_{\bar{j}}^{\prime}-\gamma_{\bar{j}}\ll\gamma_{\bar{j}}$,
one can rigorously show\cite{Fregoso13_Electrical_detection_of_TQPT} that at
the lowest order in $\delta\gamma_{\bar{j}}$, Eq. (\ref{eq:dGjj})
becomes

\begin{eqnarray*}
\Delta G_{jj}\left(0\right) & \approx&\delta\gamma_{\bar{j}}\frac{e^{2}}{h}\int_{-\infty}^{\infty}d\omega\;\left[-\frac{dn_{j}\left(\omega\right)}{d\left(eV_{j}\right)}\right]_{eV_{j}=0}\\
 & &\times\text{Tr }\left[\mathbf{t}_{ee}^{LR}\left(\mathbf{t}_{ee}^{LR}\right)^{\dagger}+\mathbf{t}_{eh}^{LR}\left(\mathbf{t}_{eh}^{LR}\right)^{\dagger}\right]_{\omega},
\end{eqnarray*}
i.e., proportional to the thermal transmission. 

\begin{figure}[t]
\begin{centering}
\includegraphics[bb=0bp 0bp 520bp 250bp,clip,scale=1]{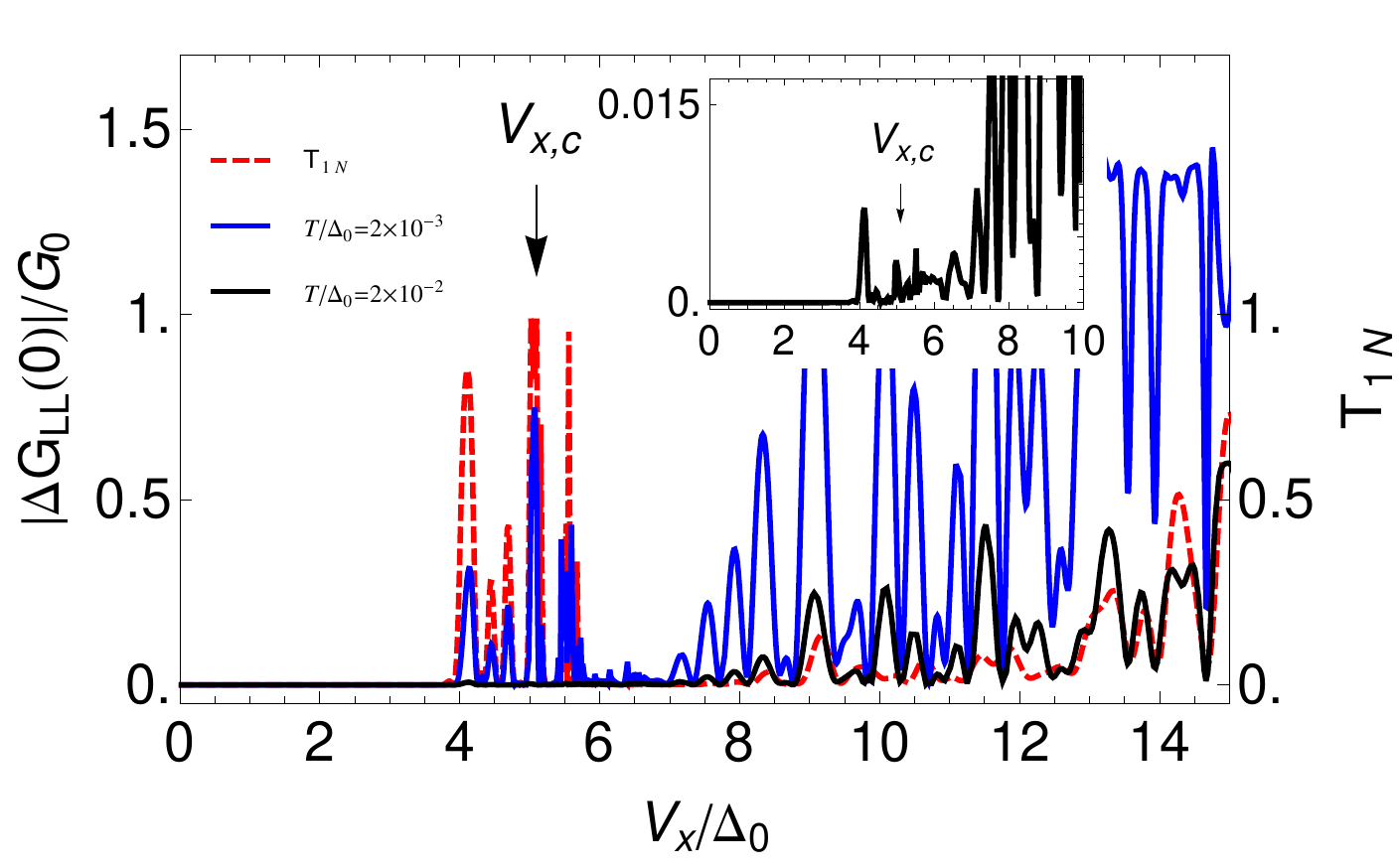}\caption{\label{fig:dGLL}Solid lines represent $\left|\Delta G_{LL}\left(0\right)\right|$
in Eq. (\ref{eq:dGjj}) as a function of $V_{x}$ at $T/\Delta_{0}=0.02$ (black line)
comparable to the experimental temperature $T_{\text{exp}}\approx60\ $mK,
 and at $T/\Delta_{0}=0.002$ (blue line). These curves
approximately follow the (thermal) transmission probability $\mathcal{T}_{1N}$
(dashed red line), which shows a maximum $\mathcal{T}_{1N}=1$ at
$V_{x,c}/\Delta_{0}\approx5.2$, corresponding to the TQPT (see arrow). All curves have been calculated with the
same configuration of the disorder potential as in Figs. \ref{fig:GLL_GRR_pdiag}
and \ref{fig:Gall}, and correspond to $\upsilon_{\mu}/\Delta_{0}=2$
{[}i.e., line of dots in Fig. \ref{fig:GLL_GRR_pdiag}(c){]}. Inset: Plot of $\left|\Delta G_{LL}\left(0\right)\right|$ at $T/\Delta_{0}=0.02$ in a smaller scale showing that, although weaker, the signal persists and is still experimentally measurable. }
\end{centering}
\end{figure}

In Fig. \ref{fig:dGLL} we show $\left|\Delta G_{LL}\left(0\right)\right|$
as a function of $V_{x}$ for the same parameters as before (see Table
\ref{tab:parametersTB}), for $\upsilon_{\mu}/\Delta_{0}=2$ and for
$\gamma_{R}=0.94\ t$ and $\gamma_{R}^{\prime}=0.01\ t$, and $\gamma_{L}=0.85\ t$.
The black solid line corresponds to the experimental temperature $T_{\text{exp}}/\Delta_{0}=0.02$,
while the blue solid line corresponds to a temperature $T/\Delta_{0}=2\times10^{-3}$
(one order of magnitude smaller). We also show a vertical cut of $\mathcal{T}_{1N}$
{[}corresponding to $\upsilon_{\mu}/\Delta_{0}=2$ in Fig. \ref{fig:GLL_GRR_pdiag}(c){]}
as a red dashed line. That curve indicates the location of the TQPT
(i.e., when$\mathcal{T}_{1N}\approx1$) in a theoretically isolated
wire, which occurs at $V_{x,c}/\Delta_{0}\approx5.2$ (indicated by
a blue dot in the horizontal axis in Fig. \ref{fig:dGLL}). For $V_{x}>V_{x,c}$,
strictly speaking the system remains in the topologically non-trivial
phase, but the strong fluctuations of $\mathcal{T}_{1N}$ indicate
a very fragile topological protection of the Majorana modes. 

\subsection{Experimental considerations}
In order to assess the experimental feasibility of our proposal, one important aspect to consider is the effect of a finite temperature.
Comparing the black and blue lines in Fig. \ref{fig:dGLL} we can see that thermal effects
dramatically decrease the magnitude of $\Delta G_{jj}\left(0\right)$,
as can be seen in the overall reduction of the signal when the temperature
increases from $T/\Delta_{0}=2\times10^{-3}$ to $T/\Delta_{0}=2\times10^{-2}$ (comparable to the experimental value $T_{\text{exp}}=60$ mK in Ref. \cite{Mourik12_Signatures_of_MF}).
While at higher fields $V_{x}>V_{x,c}$ the signal is  still clearly visible at $T/\Delta_{0}=2\times10^{-2}$, closely following the fluctuations in $\mathcal{T}_{1N}$,  near the critical field $V_{x,c}$ the signal drops to $\left|\Delta G_{jj}\left(0\right)\right|\sim3\times10^{-3}e^{2}/h$ (see inset in Fig. \ref{fig:dGLL}). Although this value is still experimentally measurable, it would be desirable to minimize thermal effects in order to have a stronger signal to detect the TQPT. 

In what follows, we show that the effect can still be measured under reasonable experimental conditions. An important point which should be taken into account to minimize thermal effect is that our proposal
is expected to work best for short wires, where the maximal ratio
$L_{\text{w}}/\xi$ is not too large (in Fig. \ref{fig:dGLL}, using
the parameters in Table \ref{tab:parametersTB}, we estimate an upper
bound $L_{\text{w}}/\xi\approx15$). Therefore, using shorter nanowires should yield more robust signals, albeit
at the cost of less resolution. This is because the visibility
of the electrical signal crucially depends on the width $~\hbar v_{F}/L_{\text{w}}$
of the peak in $\left|\Delta G_{LL}\left(0\right)\right|$ (see last paragraph in Sec. \ref{sec:intuitive} for an intuitive explanation). Therefore,
a very narrow peak $\hbar v_{F}/L_{\text{w}}\ll T$ might be hard to detect, or could be washed away
by thermal effects or other dissipative mechanisms not considered
here. Lower base temperatures or larger induced gaps should also produce a stronger signal, as can be seen in Fig. \ref{fig:dGLL} (blue line). None of these requirements represent an intrinsic experimental limitation in future samples or experiments. For instance, base temperatures of the order of $T_{\text{exp}}\approx20$ mK have been recently reported in Ref. \cite{Chang15_Hard_gap_in_SM_SC_NWs}, which would produce $T/\Delta_{0}\sim 7.10^{-3}$, allowing a stronger signal and better resolution of the transition near the critical field.

We also note that the magnetic fields required to see the signal are also within experimental reach. For the nanowires studied in Ref. \cite{Mourik12_Signatures_of_MF},  the large $g\approx 50$ factor produces $V_{x}/B\approx1.5$ meV/T. Assuming a maximal magnetic field of $B_{\text{max}}\sim 2$ T (see for instance Fig. S2 in the supplementary material in that reference) the Zeeman energy can be made as large as $V_{x}\approx3$ meV. Recalling that the experimentally induced gap is estimated in $\Delta_{0}\approx0.25$ meV, we conclude that $V_{x,\text{max}}/\Delta_{0}\sim12$, which implies that  the range of energies in Fig. \ref{fig:dGLL} is perfectly feasible.
We stress that an important requirement for this proposal is that the nanowire must
be shorter than the phase-relaxation length $L_{\text{w}}<L_{\phi}$
for the two end-MBS to hybridize coherently, a condition that is typically very well met in mesoscopic samples.

Overall, from the above discussion we conclude that $\left|\Delta G_{jj}\left(0\right)\right|$
is a bona fide indicator of the TQPT and the topological stability
of the MBS at low enough temperatures. While the experimental details will obviously depend on non-universal quantities (such as the size of the SC gap, length of the wire, degree of disorder, etc), our results in Fig. 5 indicate that the nonlocal conductance effect is definitely an experimentally observable quantity at finite temperatures. From a general point of view, our proposed experiment is much easier than either braiding or fractional Josephson effect (although harder perhaps than the straight ZBCP measurement).

Finally, although we have suggested
using the pinch off gates as a physical way to effectively tune the
coupling to the normal contacts, this is not necessarily the only
way to change the parameter $\gamma_{jj}$. For instance, schemes
using quantum dots (QDs) between the normal contact and the Majorana
nanowire\cite{Leijnse14_Parity_qubits_and_MBS,Wang14_Cross_correlations_mediated_by_MBS,Vernek14_Subtle_leakage_of_MBS_into_QD}
(i.e., N-QD-S-QD-N setups) will also serve the same purpose. In this
case, it would be relatively easy to modify the transparency of the
coupling to the lead by changing the gate voltages in each QD. However,
the QDs should be large enough to avoid strong Coulomb effects, which
might introduce unwanted effects (e.g., Kondo effect\cite{hewson})
complicating the experimental interpretation. We also mention that
in Ref. \cite{Akhmerov11_Quantized_conductance_in_disordered_wire},
an alternative method to detect the TQPT based on the measurement
of the current shot noise was proposed, which would be a complementary
to the idea discussed here.

\section{\label{sec:intuitive}Intuitive theoretical picture}

In this section we provide a simple theoretical framework to interpret
our numerical simulations in previous sections. To that end we focus
on a simplified version of the Hamiltonian $H_{\text{NW}}$ in Eqs.
(\ref{eq:H_0}) and (\ref{eq:H_SC}), which will allow us to obtain
an exact solution, therefore providing a valuable physical insight,
while retaining at the same time the relevant physics. These simplifications
will not modify our main conclusions because they do not depend on
the details of $H_{\text{NW}}$ itself, but on its \textit{symmetry
class} (i.e., class D in this case) which is a robust feature. Therefore,
for the present purposes we assume a uniform chemical potential $\mu_{0}=\delta\mu\left(x\right)=0$.
In this simplified model, disorder enters only through the inhomogeneous
pair potential $\Delta\left(x\right)$, which we now assume to be
generic and not necessarily of the form (\ref{eq:Delta_profile}). 

\begin{figure}[t]
\begin{centering}
\includegraphics[bb=40bp 220bp 260bp 350bp,clip]{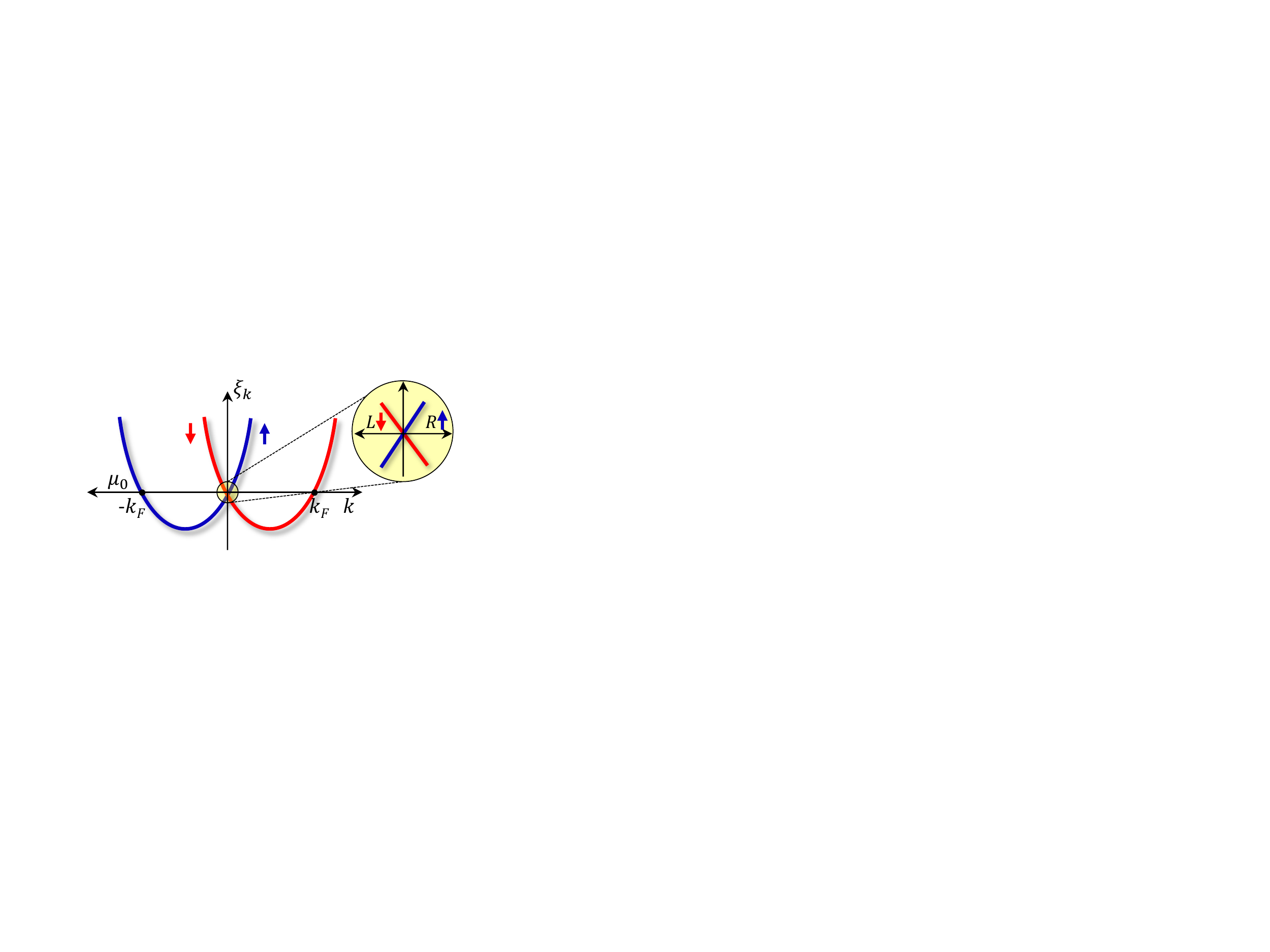}\caption{\label{fig:helical}Dispersion relation of a nanowire with Rashba
spin-orbit coupling, in the absence of proximity-induced pairing $\Delta\left(x\right)=0$,
in the absence of Zeeman field $V_{x}=0$ and for $\mu_{0}=\delta\mu\left(x\right)=0$.
Under these conditions, chosen to simplify the theoretical model in
Eq. (\ref{eq:H_helical}), a helical liquid arises near the point
$k=0$.}
\end{centering}
\end{figure}

It is simpler to start the analysis from the uniform case with periodic
boundary conditions, where the band theory helps to visualize the
relevant physics related to the TQPT occurring near the point $k=0$,
at the intersection of the spin-orbit coupled bands with different
spin projection {[}see Fig. \ref{fig:helical}{]}. The modes at finite
momentum $\pm k_{F}$ are assumed to be gapped by the SC paring interaction
(not shown in the picture), and decouple from the relevant sector
at $k=0$. Projecting the original fermionic operator around this
point and linearizing the bands results in a helical liquid model
described by the Hamiltonian\cite{Bagrets12_Class_D_spectral_peak_in_Majorana_NW}

\begin{eqnarray}
H_{\text{NW}}&= & \int dx\;\left[-i\hbar v_{F}\left(\psi_{R}^{\dagger}\partial_{x}\psi_{R}-\psi_{L}^{\dagger}\partial_{x}\psi_{L}\right)\right.\nonumber \\
 & &\left.+\Delta\left(x\right)\left(\psi_{L}\psi_{R}+\text{H.c.}\right)+\left(V_{x}\psi_{R}^{\dagger}\psi_{L}+\text{H.c.}\right)\right],\label{eq:H_helical}
\end{eqnarray}
where $\psi_{R}\simeq\psi_{\uparrow}\left(x\right)$ and $\psi_{L}\simeq\psi_{\downarrow}\left(x\right)$
result from spin-momentum locking around $k=0$ due to the spin-orbit
interaction. We now introduce the Majorana basis

\begin{eqnarray}
\boldsymbol{\eta}_{1}\left(x\right) & =  \left(\begin{array}{c}
\eta_{1,R}\\
\eta_{1,L}
\end{array}\right)\equiv\frac{1}{\sqrt{2}}\left(\begin{array}{c}
-\psi_{R}-\psi_{R}^{\dagger}\\
i\psi_{L}-i\psi_{L}^{\dagger}
\end{array}\right),\label{eq:Majo1}\\
\boldsymbol{\eta}_{2}\left(x\right) & =\left(\begin{array}{c}
\eta_{2,R}\\
\eta_{2,L}
\end{array}\right)\equiv\frac{1}{\sqrt{2}}\left(\begin{array}{c}
i\psi_{R}-i\psi_{R}^{\dagger}\\
\psi_{L}+\psi_{L}^{\dagger}
\end{array}\right),\label{eq:Majo2}
\end{eqnarray}
in terms of which (\ref{eq:H_helical}) splits into two independent
modes 
\begin{eqnarray}
H_{\text{NW}} & = &\frac{1}{2}\sum_{n=1,2}\int dx\;\boldsymbol{\eta}_{n}^{T}\left(x\right)\Biggl\{-i\hbar v_{F}\hat{\boldsymbol{\tau}}_{z}\partial_{x}\nonumber \\
 && -\left[V_{x}+\left(-1\right)^{n}\Delta\left(x\right)\right]\hat{\boldsymbol{\tau}}_{y}\Biggr\}\boldsymbol{\eta}_{n}\left(x\right),\label{eq:H_helical_majo}
\end{eqnarray}
where we have introduced the Pauli matrices $\hat{\boldsymbol{\tau}}_{i}$
acting on $LR$ space. The emergence of Majorana zero-modes can be
easily seen by solving the eigenvalue equation for $E=0$ 
\begin{eqnarray}
\left\{ -i\hbar v_{F}\hat{\boldsymbol{\tau}}_{z}\partial_{x}-\left[V_{x}+\left(-1\right)^{n}\Delta\left(x\right)\right]\hat{\boldsymbol{\tau}}_{y}\right\} \boldsymbol{\eta}_{n}\left(x\right) & = &0,\label{eq:eigen}
\end{eqnarray}
whose solution is

\begin{eqnarray}
\boldsymbol{\eta}_{n}\left(x\right) & =&\exp\left\{ \frac{1}{\hbar v_{F}}\int_{0}^{x}dx^{\prime}\ \left[V_{x}+\left(-1\right)^{n}\Delta\left(x^{\prime}\right)\right]\hat{\boldsymbol{\tau}}_{x}\right\} \boldsymbol{\eta}_{n}\left(0\right).\label{eq:solution}
\end{eqnarray}
We now define the zero-energy eigenmodes

\begin{eqnarray}
\boldsymbol{\chi}_{n}^{\pm}\left(x\right) & =\exp\left\{ \pm\frac{1}{\hbar v_{F}}\int_{0}^{x}dx^{\prime}\;\left[V_{x}+\left(-1\right)^{n}\Delta\left(x^{\prime}\right)\right]\right\} \left(\begin{array}{c}
1\\
\pm1
\end{array}\right),\label{eq:eigenmodes}
\end{eqnarray}
in terms of which the expression for a generic MBS localized at the
origin (i.e., the left end of the wire) is 

\begin{eqnarray}
\boldsymbol{\eta}\left(x\right) & =&a_{1}\boldsymbol{\chi}_{1}\left(x\right)+a_{2}\boldsymbol{\chi}_{2}\left(x\right).\label{eq:generic_majorana}
\end{eqnarray}
Although the form of the eigenmodes (\ref{eq:eigenmodes}) is more
convenient for our purposes, we mention here that one can easily bring
this expression into the more familiar form of the Jackiw-Rebbi soliton
solution\cite{Jackiw76_Jackiw_Rebbi_soliton} applying a rotation
along the $\hat{y}-$axis $\hat{R}=e^{i\frac{\pi}{4}\hat{\boldsymbol{\tau}}_{y}}$,
which transforms to the usual eigenvectors of the operator $\hat{\boldsymbol{\tau}}_{z}$.
In order to ensure the existence of MBS we need to find normalizable
solutions that decrease sufficiently fast as $x\rightarrow\infty$
and that satisfy generic boundary conditions. For a wire of length
$L_{\text{w}}$, we can define the quantity

\begin{eqnarray}
\lambda_{n} & =&\frac{1}{L_{\text{w}}\hbar v_{F}}\int_{0}^{L_{\text{w}}}dx^{\prime}\;\left[V_{x}+\left(-1\right)^{n}\Delta\left(x^{\prime}\right)\right],\label{eq:lambda_j}
\end{eqnarray}
which in the limit $L_{\text{w}}\rightarrow\infty$ corresponds to
the Lyapunov exponent of the system at zero energy for the channel
$n$. In terms of these quantities, note that there are two possible
situations:\cite{kitaev2001} 
\begin{enumerate}
\item Both $\lambda_{1}$ and $\lambda_{2}$ have the same sign, in which
case we need to choose either $\boldsymbol{\eta}\left(x\right)=a_{1}\boldsymbol{\chi}_{1}^{+}\left(x\right)+a_{2}\boldsymbol{\chi}_{2}^{+}\left(x\right)$
or $\boldsymbol{\eta}\left(x\right)=a_{1}\boldsymbol{\chi}_{1}^{-}\left(x\right)+a_{2}\boldsymbol{\chi}_{2}^{-}\left(x\right)$
in Eq. (\ref{eq:generic_majorana}), the sign depending on which of
the modes decays for $x>0$. Since there are two decaying contributions
allowed, we can satisfy generic boundary conditions at the origin.
For instance, if the system is a trivial insulator for $x<0$, then
the boundary condition $\boldsymbol{\eta}\left(0\right)=\left(0,0\right)^{T}$
must be imposed. This is verified with $a_{1}+a_{2}=0$. Other boundary
conditions for open wires will be analyzed later. 
\item The Lyapunov exponents $\lambda_{1}$ and $\lambda_{2}$ have different
signs. Then Eq. (\ref{eq:generic_majorana}) is a linear combination
of spinors $\boldsymbol{\chi}^{+}$ and $\boldsymbol{\chi}^{-}$.
This makes it impossible to satisfy generic boundary conditions, except
for accidental situations which are not protected against local perturbations.
For instance, in our previous example of a vanishing boundary condition
at the origin, the condition $\boldsymbol{\eta}\left(0\right)=\left(0,0\right)^{T}$
implies that $a_{1}+a_{2}=a_{1}-a_{2}=0$, which can only be satisfied
for $a_{1}=a_{2}=0$. Therefore the MBS does not exist.
\end{enumerate}
From this analysis, we conclude that the TQPT occurs when one of the
Lyapunov exponents $\lambda_{n}$ passes through zero and changes
sign, making explicit the connection between the localization properties
of a D-class nanowire and its topological properties. This is a robust
feature which is independent of the details of the microscopic Hamiltonian
as it depends only on the symmetry classification. Assuming that the
magnetic field $V_{x}$ is such that $\lambda_{2}>0$, the condition
for the topological phase reduces to 

\begin{eqnarray}
\lambda_{1} & =&\frac{1}{\hbar v_{F}}\left[V_{x}-\bar{\Delta}\right]>0,\label{eq:lamba1}
\end{eqnarray}
where we have defined the average gap $\bar{\Delta}\equiv\frac{1}{L_{\text{w}}}\int_{0}^{L_{\text{w}}}dx^{\prime}\;\Delta\left(x^{\prime}\right)$.
Note that this expression coincides with the expression derived by
Sau \textit{et al.} $V_{x}>\sqrt{\mu_{0}^{2}+\Delta_{0}^{2}}$ for
the ideal system, i.e., for a uniform $\Delta\left(x\right)\rightarrow\Delta_{0}$
and for $\mu_{0}=0$.\cite{Sau10_Proposal_for_MF_in_semiconductor_heterojunction}

\begin{figure}[t]
\begin{centering}
\includegraphics[bb=40bp 20bp 670bp 400bp,clip,scale=0.37]{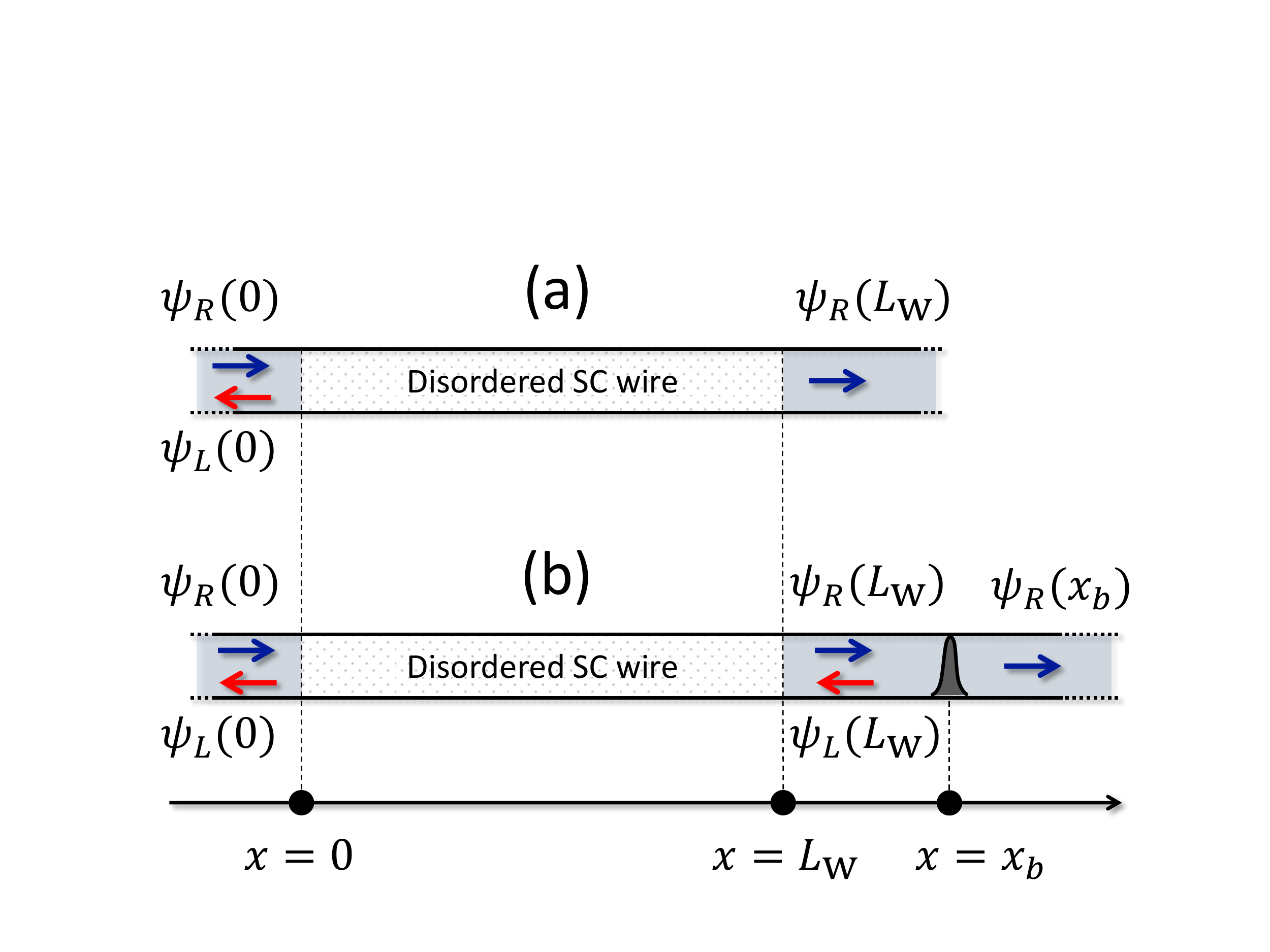}\caption{Schematic view of the scattering across a disordered Majorana wire
in the NSN configuration. In (a) a reflectionless boundary at the
right end is assumed and, consequently, the boundary condition $\boldsymbol{\psi}_{L}\left(L_{\text{w}}\right)=\left(0,0\right)^{T}$
must be imposed. This is a very special case, typically incompatible
with the experimental situation. In (b) we assume a generic barrier
$V_{b}\left(x\right)$ at the right end inducing a reflection amplitude
matrix $\mathbf{r}_{b}\neq0$. In this case different boundary conditions
must be imposed {[}see Eq. (\ref{eq:scattering_matrix_barrier}){]}.\label{fig:open_wire}}
\end{centering}\end{figure}

\subsection{Open wires}

We now assume that our wire is connected to conducting leads at both
ends, and focus on the transport across the NSN configuration at zero
energy, as depicted in Fig. \ref{fig:open_wire}(a). We define the
scattering matrix \cite{Akhmerov11_Quantized_conductance_in_disordered_wire} 

\begin{eqnarray}
\left(\begin{array}{c}
\boldsymbol{\psi}_{R}\left(L_{\text{w}}\right)\\
\boldsymbol{\psi}_{L}\left(0\right)
\end{array}\right) & =\mathcal{S}_{0}\left(\begin{array}{c}
\boldsymbol{\psi}_{R}\left(0\right)\\
\boldsymbol{\psi}_{L}\left(L_{\text{w}}\right)
\end{array}\right),\qquad\mathcal{S}_{0}=\left(\begin{array}{cc}
\mathbf{t}_{0} & \mathbf{r}_{0}^{\prime}\\
\mathbf{r}_{0} & \mathbf{t}_{0}^{\prime}
\end{array}\right),\label{eq:scattering_matrix_wire}
\end{eqnarray}
where $\boldsymbol{\psi}_{\nu}\left(x\right)\equiv\left(\eta_{1,\nu}\left(x\right),\ \eta_{2,\nu}\left(x\right)\right)^{T}$
are $\nu-$moving (with $\nu=\left\{ L,R\right\} $) scattering Majorana
states in the left and right leads ($x=0$ and $x=L_{\text{w}}$,
respectively). In the Majorana basis, $\mathcal{S}_{0}$ is a real
orthonormal matrix $\mathcal{S}_{0}^{T}=\mathcal{S}_{0}^{\dagger}=\mathcal{S}_{0}^{-1}$.\cite{Akhmerov11_Quantized_conductance_in_disordered_wire}
Since from Eq. (\ref{eq:H_helical}), the modes $n=\left\{ 1,2\right\} $
are decoupled and independent, the transmission and reflection matrices,
$\mathbf{t}_{0},\mathbf{t}_{0}^{\prime}$ and $\mathbf{r}_{0},\mathbf{r}_{0}^{\prime}$
respectively, acquire a diagonal form, an can be diagonalized independently
with diagonal elements $t_{0,n},t_{0,n}^{\prime}$ and $r_{0,n},r_{0,n}^{\prime}$.
Without loss of generality, in what follows we assume that the only
incident modes are right-moving modes arriving from the left lead.
This imposes the boundary conditions {[}see Fig. \ref{fig:open_wire}(a){]}

\begin{eqnarray}
\boldsymbol{\psi}_{L}\left(L_{\text{w}}\right)=\left(\begin{array}{c}
0\\
0
\end{array}\right) & \quad\quad & \boldsymbol{\psi}_{R}\left(0\right)=\left(\begin{array}{c}
1\\
1
\end{array}\right),\label{eq:reflectionless_bc}
\end{eqnarray}
which in combination with Eqs. (\ref{eq:solution}) and (\ref{eq:scattering_matrix_wire}),
allow to obtain closed analytical expressions for the reflection and
transmission coefficients %

\begin{eqnarray}
t_{0,n} & =&\cosh^{-1}\left(L_{\text{w}}\lambda_{n}\right),\label{eq:tj_a}\\
r_{0,n} & =&-\tanh\left(L_{\text{w}}\lambda_{n}\right),\label{eq:rj_a}
\end{eqnarray}
and we recover Eq. (\ref{eq:Tn}) for the transmission probability.

Exactly at the TQPT, the determinant of the reflection matrix vanishes
and a ``Majorana channel'' with perfect transmission opens at zero
energy. As mentioned in Sec. \ref{sec:transfer}, Akhmerov \textit{et
al}.\cite{Akhmerov11_Quantized_conductance_in_disordered_wire} derived
a suitable topological invariant for a dirty class D nanowire directly
in terms of the reflection matrix as $Q=\text{sign Det }\mathbf{r}_{0}=\text{sign Det }\mathbf{r}_{0}^{\prime}=\prod_{n}\tanh\lambda_{0,n}$,
and suggested that the TQPT could be observed as a quantized peak
in the thermal conductance through the nanowire $G_{\text{th}}/G_{0}=\text{Tr }\left(\mathbf{t}_{0}\mathbf{t}_{0}^{\dagger}\right)=\sum_{n}\cosh^{-2}\left(L_{\text{w}}\lambda_{n}\right)$,
where we recover the result in Sec. \ref{sec:transfer}. This is consistent
with the results in Ref. \cite{Motrunich01_Disorder_in_topological_1D_SC},
where the authors predicted that the transition from topologically
trivial to topologically non-trivial phases should be a \textit{delocalization}
transition, and at both sides of this point the system should be generically
localized at zero energy. However, unfortunately a Majorana channel
is necessarily neutral (i.e., particles and holes have equal weight
in the MBS wavefunction) and therefore cannot support an electrical
current. On the other hand, direct thermal transport measurements
could provide evidence of the transition,\cite{Akhmerov11_Quantized_conductance_in_disordered_wire}
but this remains an experimental challenge.%

To understand better our experimental proposal in Sec. \ref{sec:pdiag}
we first note that the form of Eqs. (\ref{eq:tj_a}) and (\ref{eq:rj_a})
are a consequence of the particular ``reflectionless'' boundary
conditions (\ref{eq:reflectionless_bc}) at the right end. In other
words, in Fig. \ref{fig:open_wire}(a) at the right end of the wire,
the barrier at $x=L_{\text{w}}$ is ``transparent'', and all right-moving
Majorana states $\boldsymbol{\psi}_{R}\left(L_{\text{w}}\right)$
that are transmitted to the right-end of the nanowire disappears in
the right lead. As shown in Ref. \cite{Fregoso13_Electrical_detection_of_TQPT},
this is not the most general situation. The generic presence of a
barrier $V_{b}\left(x\right)$ at the end of the nanowire induces
some probability of reflection, and imposes a non-vanishing amplitude
$\boldsymbol{\psi}_{L}\left(L_{\text{w}}\right)$ {[}see Fig. \ref{fig:open_wire}(b){]}.
More physically, any potential profile at the end nanowire, or the
presence of pinch off gates could play the role of a barrier inducing
a non-ideal coupling to the right-lead. For simplicity, let us consider
a point-like scatterer sitting at some point $x_{b}>L_{\text{w}},$
as depicted in Fig. \ref{fig:open_wire}(b). The crucial point is
that, in the presence of this new barrier, the reflectionless boundary
conditions (\ref{eq:reflectionless_bc}) are no longer possible. Assuming
that the potential barrier $V_{b}\left(x\right)$ induces reflection
and transmission amplitudes, $r_{b,n}$ and $t_{b,n}$ respectively
(subject to the unitary condition $\left|r_{b,n}\right|^{2}+\left|t_{b,n}\right|^{2}=1$),
the scattering matrix obeys
\begin{eqnarray}
\left(\begin{array}{c}
\boldsymbol{\psi}_{R}\left(x_{b}\right)\\
\boldsymbol{\psi}_{L}\left(L_{\text{w}}\right)
\end{array}\right)=\mathcal{S}_{b}\left(\begin{array}{c}
\boldsymbol{\psi}_{R}\left(L_{\text{w}}\right)\\
\boldsymbol{\psi}_{L}\left(x_{b}\right)
\end{array}\right),\qquad\mathcal{S}_{b}=\left(\begin{array}{cc}
\mathbf{t}_{b} & \mathbf{r}_{b}^{\prime}\\
\mathbf{r}_{b} & \mathbf{t}_{b}^{\prime}
\end{array}\right),\label{eq:scattering_matrix_barrier}
\end{eqnarray}
and the new boundary conditions for right-moving Majorana states arriving
from the left lead are
\begin{eqnarray}
\boldsymbol{\psi}_{L}\left(x_{b}\right)=\left(\begin{array}{c}
0\\
0
\end{array}\right) & \quad\quad\boldsymbol{\psi}_{R}\left(0\right)=\left(\begin{array}{c}
1\\
1
\end{array}\right) & .\label{eq:boundary_condition_scatterer}
\end{eqnarray}
In combination with Eqs. (\ref{eq:scattering_matrix_wire}) and (\ref{eq:scattering_matrix_barrier}),
we obtain the transmission and reflection amplitudes at the left-end
of the \textit{complete} system (nanowire \textit{and} barrier):%

\begin{eqnarray}
t_{n} & =\frac{t_{b,n}t_{0,n}}{1-r_{b,n}r_{0,n}},\label{eq:tj_total}\\
r_{n} & =r_{0,n}+t_{0,n}\left(\frac{r_{b,n}}{1-r_{b,n}r_{0,n}}\right)t_{0,n}.\label{eq:rj_total}
\end{eqnarray}
In particular, the last term in Eq. (\ref{eq:rj_total}) physically
represents processes in which the right-moving Majorana mode is transmitted
to the right-end of the nanowire with amplitude $t_{0,n}$ and is
reflected back as a left-mover with amplitude $r_{b,n}$. The denominator
in $r_{b,n}\left(1-r_{b,n}r_{0,n}\right)^{-1}=r_{b,n}+r_{b,n}r_{0,n}r_{b,n}+\dots$,
represents an infinite sum of all backward and forward internal reflection
processes occurring in the wire. Importantly, even though $r_{n}$
is a local quantity involving the reflection at the left-end of the
nanowire, Eq. (\ref{eq:rj_total}) explicitly contains non-local contributions
involving scattering at the right-end, and its form is closely related
to Eqs. (\ref{eq:GLL_final_symm-1}) and (\ref{eq:GRR_final_symm-1})
for the local differential conductances. This is a milestone result
in phase-coherent mesoscopic transport which has been well-known for
almost thirty years.\cite{Buttiker86_Landauer_Buttiker_paper} %

The above considerations summarize the main theoretical ideas in this
work. Assuming that $r_{b,n}$ is a parameter that can be modified
\textit{in situ} in the experiment (as is the case of the pinch off
gates in Ref. \cite{Mourik12_Signatures_of_MF}), Eq. (\ref{eq:rj_total})
shows that a small variation $\delta r_{b,n}$ gives rise to a modification
$\delta r_{n}\propto\cosh^{-2}\left(L_{\text{w}}\lambda_{n}\right)\delta r_{b,n}$,
which would be non-vanishing precisely at the TQPT and which could
be detected in \textit{electrical} measurements. This is the main
idea of our proposal, and the main reason for us to propose experiments
in the NSN geometry in order to establish the existence of the TQPT
and the MBS in Majorana nanowires.

\section{\label{sec:summary}Summary and conclusions}

We have explored the transport properties of disordered Majorana nanowires
in the NSN configuration with the nanowire being the superconducting
S part and the two N parts are ordinary metallic tunneling contacts
with suitable gates controlling their tunnel barriers. This type of
geometry is being explored at present by experimental groups studying
Majorana bound states, and consequently, our study might be of relevance
for the interpretation of these results. The NSN configuration allows
to access qualitatively new information about the topological properties
of the system through a direct study of non-local correlations inherent
in the MBS which cannot be done in the NS geometry mostly used in
the experimental Majorana measurements so far. Physically, this is
possible because in the NSN configuration one can test the \textit{bulk}
properties, in addition to the boundary properties which are the only
properties accessible in NS contacts. In our work we have adopted
a comprehensive point of view, which links the deep theoretical aspects
(i.e., topological invariants, topological classification, topological
quantum phase transition and topological phase diagram) with the experimental
observables (i.e., tunneling transport). We have also proposed a useful
tool, i.e., the difference of local conductances Eq. (\ref{eq:dGjj}),
to detect the TQPT occurring as a function of the applied Zeeman field
and to assess the topological protection of a given system experimentally.
The experimental signal Eq. (\ref{eq:dGjj}) is expected to be stronger
and more robust to thermal broadening effects for ``short'' wires
with ratio $L_{\text{w}}/\xi$ not too large ($L_{\text{w}}/\xi\approx15$
in this work) and $L_{\text{w}}<L_{\phi}$, i.e., smaller than the
phase-relaxation length. We stress that this proposal to detect the
TQPT is qualitatively different from the study of non-local correlations
in the shot noise measurements.\cite{Bolech07_Shot_noise_Majorana,Nilsson_PRL08,Liu12_Current_noise_correlation,Zocher13_Current_cross_correlations_in_Majorana_NW}
Despite the simplifications assumed in this work, we note that our
main ideas do not rely on the details of our model, but on generic
symmetry properties of class D Bogoliubov-de Gennes Hamiltonians.
In particular, the fact that the TQPT correspond to a delocalized
point at zero energy is a robust feature in these non-interacting
Hamiltonians. In the presence of interactions the theoretical description
of transport becomes much harder and remains an open issue. However,
we speculate that the main idea behind Eq. (\ref{eq:dGjj}) should
remain valid in that case too. Interestingly, using the framework
of Abelian bosonization, in Ref. \cite{Lobos12_Interplay_disorder_interaction_Majorana_wire}
it was shown that the low-temperature properties of a disordered class
D wire with repulsive short-range electron-electron interactions (i.e.,
dimensionless Luttinger parameter\cite{Giamarchi_book} $K<1$) are
adiabatically connected to those of a non-interacting wire (i.e.,
with $K=1$), provided the system remains in the topological phase
as the interaction is adiabatically ``turned on''. In particular,
the delocalized nature of the TQPT in the interacting case can be
inferred using an instanton calculation in the presence of disorder,
where the equivalent of the localization length (i.e., the exponent
of the instanton action) diverges at the critical point.\cite{Lobos12_Interplay_disorder_interaction_Majorana_wire} 
\\
This work is supported by Microsoft Q, LPS-CMTC and JQI-NSF-PFC. AML acknowledges
useful discussions with James Williams, L. Rokhinson and A. Akhmerov.
\appendix

\section{\label{sec:appendix}Derivation of Eqs. (\ref{eq:GLL_final_symm-1})-(\ref{eq:GRR_final_symm-1})}

Starting from Eqs. (\ref{eq:Hw}) and (\ref{eq:H_mix}), the expression
for the electric current flowing through the contacts is $I_{j}=e\langle dN_{j}/dt\rangle=ie\langle[H,N_{j}]\rangle/\hbar=ie\langle[H_{\text{mix}},N_{j}]\rangle/\hbar$,
which can be written in terms of the Green's function at the contacts
\cite{Cuevas96_Hamiltonian_approach_to_SC_contacts,meir92} 

\begin{eqnarray}
I_{L} & =&\frac{ie}{\hbar}\sum_{\sigma}t_{L}\left[\left\langle d_{L,\sigma}^{\dagger}c_{1,\sigma}\right\rangle -\left\langle c_{1,\sigma}^{\dagger}d_{L,\sigma}\right\rangle \right],\label{eq:currentL}\\
I_{R} & =&\frac{ie}{\hbar}\sum_{\sigma}t_{R}\left[\left\langle d_{R,\sigma}^{\dagger}c_{N,\sigma}\right\rangle -\left\langle c_{N,\sigma}^{\dagger}d_{R,\sigma}\right\rangle \right],\label{eq:currentR}
\end{eqnarray}
where we have defined $d_{j,\sigma}=\frac{1}{\sqrt{\mathcal{N}_{j}}}\sum_{k}d_{jk,\sigma}$
, with $j=\left\{ L,R\right\} $, and where $\mathcal{N}_{j}$ is
the number of sites in the lead $j$. With these definitions, note
that the currents are positive if particles move into the leads (i.e.,
exit the SC), and negative otherwise. On the other hand, charge conservation
demands that $I_{L}+I_{R}+I_{S}=0$, where $I_{S}$ is the excess
current that flows to ground through the SC (see Fig. \ref{fig:NSN_circuit-1}).
Within the Baym-Kadanoff-Keldysh formalism \cite{Kadanoff1989,mahan}
we define the lesser and bigger Green's functions

\begin{eqnarray}
g_{i\sigma,j\sigma^{\prime}}^{<}\left(t\right) & \equiv ie\left\langle c_{i,\sigma}^{\dagger}c_{j,\sigma}\left(t\right)\right\rangle ,\\
g_{i\sigma,j\sigma^{\prime}}^{>}\left(t\right) & \equiv-ie\left\langle c_{i,\sigma}\left(t\right)c_{j,\sigma}^{\dagger}\right\rangle ,
\end{eqnarray}
 so that we can write the currents as 
\begin{eqnarray}
I_{L} & = &\frac{e}{\hbar}t_{L}\sum_{\sigma}\int_{-\infty}^{\infty}\frac{d\omega}{2\pi}\left[g_{L\sigma,1\sigma}^{<}\left(\omega\right)-g_{1\sigma,L\sigma}^{<}\left(\omega\right)\right],\label{eq:currentL_2}\\
I_{R} & =&\frac{e}{\hbar}t_{R}\sum_{\sigma}\int_{-\infty}^{\infty}\frac{d\omega}{2\pi}\left[g_{R\sigma,N\sigma}^{<}\left(\omega\right)-g_{N\sigma,R\sigma}^{<}\left(\omega\right)\right].\label{eq:currentR_2}
\end{eqnarray}
Using equations of motion, we can express Eqs. (\ref{eq:currentL_2})
and (\ref{eq:currentR_2}) in terms of local Green's functions as
\cite{Cuevas96_Hamiltonian_approach_to_SC_contacts,meir92}

\begin{eqnarray}
I_{L} & =&-\frac{e}{h}t_{L}^{2}\sum_{\sigma}\int_{-\infty}^{\infty}d\omega\left[g_{L\sigma,L\sigma}^{0,<}\left(\omega\right)g_{1\sigma,1\sigma}^{>}\left(\omega\right)\right.\nonumber \\
 & &\left.-g_{L\sigma,L\sigma}^{0,>}\left(\omega\right)g_{1\sigma,1\sigma}^{<}\left(\omega\right)\right],\label{eq:currentL_3}\\
I_{R} & =&-\frac{e}{h}t_{R}^{2}\sum_{\sigma}\int_{-\infty}^{\infty}d\omega\left[g_{R\sigma,R\sigma}^{0,<}\left(\omega\right)g_{N\sigma,N\sigma}^{>}\left(\omega\right)\right.\nonumber \\
 & &\left.-g_{R\sigma,R\sigma}^{0,>}\left(\omega\right)g_{N\sigma,N\sigma}^{<}\left(\omega\right)\right].\label{eq:currentR_3}
\end{eqnarray}
The unperturbed Green's functions $g_{j\sigma,j\sigma}^{0,\gtrless}\left(\omega\right)$
in the leads

\begin{eqnarray}
g_{j\sigma,j\sigma}^{0,<}\left(\omega\right) & =&2\pi i\rho_{j,\sigma}^{0}\left(\omega\right)n_{j}\left(\omega\right),\label{eq:g0_lesser}\\
g_{j\sigma,j\sigma}^{0,>}\left(\omega\right) & =&2\pi i\rho_{j,\sigma}^{0}\left(\omega\right)\left[n_{j}\left(\omega\right)-1\right],\label{eq:g0_bigger}
\end{eqnarray}
contain the information about the Fermi distribution functions $n_{j}\left(\omega\right)=n_{F}\left(\omega+\mu_{j}\right)$
at the leads. Substituting Eqs. (\ref{eq:g0_lesser}) and (\ref{eq:g0_bigger})
into Eqs. (\ref{eq:currentL_3}) and (\ref{eq:currentR_3}) yields
\begin{eqnarray}
\nonumber \\
I_{L} & =-\frac{ie}{h}2\pi t_{L}^{2}\sum_{\sigma}\int_{-\infty}^{\infty}d\omega\;\rho_{L,\sigma}^{0}\left(\omega\right)\left\{ n_{L}\left(\omega\right)\left[g_{1\sigma,1\sigma}^{r}\left(\omega\right)-g_{1\sigma,1\sigma}^{a}\left(\omega\right)\right]+g_{1\sigma,1\sigma}^{<}\left(\omega\right)\right\} ,\label{eq:currentL_4}\\
I_{R} & =-\frac{ie}{h}2\pi t_{R}^{2}\sum_{\sigma}\int_{-\infty}^{\infty}d\omega\;\rho_{R,\sigma}^{0}\left(\omega\right)\left\{ n_{R}\left(\omega\right)\left[g_{N\sigma,N\sigma}^{r}\left(\omega\right)-g_{N\sigma,N\sigma}^{a}\left(\omega\right)\right]+g_{N\sigma,N\sigma}^{<}\left(\omega\right)\right\} ,\label{eq:currentR_4}
\end{eqnarray}
where we have used the identity $g^{>}\left(\omega\right)-g^{<}\left(\omega\right)=g^{r}\left(\omega\right)-g^{a}\left(\omega\right)$.\cite{Kadanoff1989,mahan}
Obtaining an explicit expression for the currents $I_{L}$ and $I_{R}$
in the general case is quite cumbersome. However, since we will be
interested only in the conductance, we note that there is a great
simplification if we compute directly the conductance matrix by deriving
the currents with respect to the voltages $V_{L},V_{R}$. Then

\begin{eqnarray}
G_{LL}\equiv\frac{dI_{L}}{dV_{L}} & =&-\frac{ie^{2}}{h}2\pi t_{L}^{2}\sum_{\sigma}\int_{-\infty}^{\infty}d\omega\;\rho_{L,\sigma}^{0}\left(\omega\right)\nonumber \\ &&\times \left\{ \frac{dn_{L}\left(\omega\right)}{d\left(eV_{L}\right)}\left[g_{1\sigma,1\sigma}^{r}\left(\omega\right)-g_{1\sigma,1\sigma}^{a}\left(\omega\right)\right]+\frac{dg_{1\sigma,1\sigma}^{<}\left(\omega\right)}{d\left(eV_{L}\right)}\right\} ,\label{eq:G_LL}\\
G_{LR}\equiv\frac{dI_{L}}{dV_{R}} & =&-\frac{ie^{2}}{h}2\pi t_{L}^{2}\sum_{\sigma}\int_{-\infty}^{\infty}d\omega\;\rho_{L,\sigma}^{0}\left(\omega\right)\frac{dg_{1\sigma,1\sigma}^{<}\left(\omega\right)}{d\left(eV_{R}\right)},\label{eq:G_LR}\\
G_{RL}\equiv\frac{dI_{R}}{dV_{L}} & =&-\frac{ie^{2}}{h}2\pi t_{R}^{2}\sum_{\sigma}\int_{-\infty}^{\infty}d\omega\;\rho_{R,\sigma}^{0}\left(\omega\right)\frac{dg_{N\sigma,N\sigma}^{<}\left(\omega\right)}{d\left(eV_{L}\right)},\label{eq:G_RL}\\
G_{RR}\equiv\frac{dI_{R}}{dV_{R}} & =&-\frac{ie^{2}}{h}2\pi t_{R}^{2}\sum_{\sigma}\int_{-\infty}^{\infty}d\omega\;\rho_{R,\sigma}^{0}\left(\omega\right) \nonumber \\ &&\times\left\{ \frac{dn_{R}\left(\omega\right)}{d\left(eV_{R}\right)}\left[g_{N\sigma,N\sigma}^{r}\left(\omega\right)-g_{N\sigma,N\sigma}^{a}\left(\omega\right)\right]+\frac{dg_{N\sigma,N\sigma}^{<}\left(\omega\right)}{d\left(eV_{R}\right)}\right\} .\label{eq:G_RR}
\end{eqnarray}
Therefore, we see that the problem is reduced to finding the Green's
functions in the superconducting system. In a non-interacting system,
the full Green's function verifies the Dyson's equation in Nambu space
\cite{Cuevas96_Hamiltonian_approach_to_SC_contacts}

\begin{eqnarray}
\boldsymbol{\mathcal{G}}^{\gtrless}\left(\omega\right) & =&\left[\mathbf{1}+\boldsymbol{\mathcal{G}}^{r}\left(\omega\right)\left(\boldsymbol{\mathcal{T}}_{L}+\boldsymbol{\mathcal{T}}_{R}\right)\right]\boldsymbol{\mathcal{G}}^{0,\gtrless}\left(\omega\right)\left[\mathbf{1}+\left(\boldsymbol{\mathcal{T}}_{L}+\boldsymbol{\mathcal{T}}_{R}\right)\boldsymbol{\mathcal{G}}^{a}\left(\omega\right)\right],\label{eq:G_Dyson_Keldysh}\\
\boldsymbol{\mathcal{G}}^{\left(r,a\right)}\left(\omega\right) & =&\boldsymbol{\mathcal{G}}^{0,\left(r,a\right)}\left(\omega\right)+\boldsymbol{\mathcal{G}}^{0,\left(r,a\right)}\left(\omega\right)\left(\boldsymbol{\mathcal{T}}_{L}+\boldsymbol{\mathcal{T}}_{R}\right)\boldsymbol{\mathcal{G}}^{\left(r,a\right)}\left(\omega\right),\label{eq:G_Dyson_ra}
\end{eqnarray}
where we have introduced the Nambu notation 
\begin{eqnarray}
\boldsymbol{\mathcal{G}}_{i\sigma,j\sigma^{\prime}}^{\nu}\left(z\right) & =&\left(\begin{array}{cc}
g_{i\sigma,j\sigma^{\prime}}^{\nu}\left(z\right) & f_{i\sigma,j\sigma^{\prime}}^{\nu}\left(z\right)\\
\bar{f}_{i\sigma,j\sigma^{\prime}}^{\nu}\left(z\right) & \bar{g}_{i\sigma,j\sigma^{\prime}}^{\nu}\left(z\right)
\end{array}\right),
\end{eqnarray}
 with $\nu=\left\{ >,<,r,a\right\} $, and where 
\begin{eqnarray}
\boldsymbol{\mathcal{T}} & _{j}=\left(\begin{array}{cc}
t_{j} & 0\\
0 & -t_{j}
\end{array}\right).
\end{eqnarray}
 The unperturbed Keldysh Green's functions (i.e., computed for $t_{L}=t_{R}=0$)
are

\begin{eqnarray}
\boldsymbol{\mathcal{G}}_{i\sigma,j\sigma^{\prime}}^{0,<}\left(\omega\right) & =&2\pi i\boldsymbol{\rho}_{i\sigma,j\sigma^{\prime}}^{0}\left(\omega\right)n_{F}\left(\omega\right),\label{eq:G0_lesser}\\
\boldsymbol{\mathcal{G}}_{i\sigma,j\sigma^{\prime}}^{0,>}\left(\omega\right) & =&2\pi i\boldsymbol{\rho}_{i\sigma,j\sigma^{\prime}}^{0}\left(\omega\right)\left[n_{F}\left(\omega\right)-1\right],\label{eq:G0_bigger}\\
\boldsymbol{\rho}_{i\sigma,j\sigma^{\prime}}^{0}\left(\omega\right) & =&-\frac{1}{\pi}\text{Im}\left[\boldsymbol{\mathcal{G}}_{i\sigma,j\sigma^{\prime}}^{0,r}\left(\omega\right)\right]=\left(\begin{array}{cc}
\rho_{i\sigma,j\sigma^{\prime}}^{0}\left(\omega\right) & \zeta_{i\sigma,j\sigma^{\prime}}^{0}\left(\omega\right)\\
\zeta_{i\sigma,j\sigma^{\prime}}^{0}\left(\omega\right) & \bar{\rho}_{i\sigma,j\sigma^{\prime}}^{0}\left(\omega\right)
\end{array}\right),\label{eq:Rho}
\end{eqnarray}
We only need the derivative with respect to the voltages, which are
only in the leads. Therefore%

\begin{eqnarray*}
\frac{dg_{1\sigma,1\sigma}^{\gtrless}}{d\left(eV_{L}\right)} & =2\pi it_{L}^{2}\sum_{s}\left[\frac{dn_{L}}{d\left(eV_{L}\right)}\rho_{L}^{0}g_{1\sigma,1s}^{r}g_{1s,1\sigma}^{a}+\frac{d\bar{n}_{L}}{d\left(eV_{L}\right)}\bar{\rho}_{L}^{0}f_{1\sigma,1s}^{r}\bar{f}_{1s,1\sigma}^{a}\right],\\
\frac{dg_{1\sigma,1\sigma}^{\gtrless}}{d\left(eV_{R}\right)} & =2\pi it_{R}^{2}\sum_{s}\left[\frac{dn_{R}}{d\left(eV_{R}\right)}\rho_{R}^{0}g_{1\sigma,Ns}^{r}g_{Ns,1\sigma}^{a}+\frac{d\bar{n}_{R}}{d\left(eV_{R}\right)}\bar{\rho}_{R}^{0}f_{1\sigma,Ns}^{r}\bar{f}_{Ns,1\sigma}^{a}\right],\\
\frac{dg_{N\sigma,N\sigma}^{\gtrless}}{d\left(eV_{L}\right)} & =2\pi it_{L}^{2}\sum_{s}\left[\frac{dn_{L}}{d\left(eV_{L}\right)}\rho_{L}^{0}g_{N\sigma,1s}^{r}g_{1s,N\sigma}^{a}+\frac{d\bar{n}_{L}}{d\left(eV_{L}\right)}\bar{\rho}_{L}^{0}f_{N\sigma,1s}^{r}\bar{f}_{1s,N\sigma}^{a}\right],\\
\frac{dg_{N\sigma,N\sigma}^{\gtrless}}{d\left(eV_{R}\right)} & =2\pi it_{R}^{2}\sum_{s}\left[\frac{dn_{R}}{d\left(eV_{R}\right)}\rho_{R}^{0}g_{N\sigma,Ns}^{r}g_{Ns,N\sigma}^{a}+\frac{d\bar{n}_{R}}{d\left(eV_{R}\right)}\bar{\rho}_{R}^{0}f_{N\sigma,Ns}^{r}\bar{f}_{Ns,N\sigma}^{a}\right].
\end{eqnarray*}
Replacing these expressions into Eqs. (\ref{eq:G_LL})-(\ref{eq:G_RR}),
and using the result $g_{j\sigma,j\sigma}^{r}\left(\omega\right)-g_{j\sigma,j\sigma}^{a}\left(\omega\right)=-2\pi i\rho_{j\sigma}\left(\omega\right)$,
where we have defined the local density of states $\rho_{j\sigma}\left(\omega\right)\equiv\rho_{j\sigma,j\sigma}\left(\omega\right)$,
yields

\begin{eqnarray}
G_{LL} & = & -\frac{e^{2}}{h}\sum_{\sigma}\int_{-\infty}^{\infty}d\omega\;\gamma_{L}\left(\omega\right)\left[\frac{dn_{L}}{d\left(eV_{L}\right)}2\pi\rho_{1\sigma}-\sum_{s}\frac{dn_{L}}{d\left(eV_{L}\right)}\gamma_{L}g_{1\sigma,1s}^{r}g_{1s,1\sigma}^{a} \right. \nonumber \\ &&- \left. \sum_{s}\frac{d\bar{n}_{L}}{d\left(eV_{L}\right)}\bar{\gamma}_{L}f_{1\sigma,1s}^{r}\bar{f}_{1s,1\sigma}^{a}\right]_{\omega},\label{eq:GLL_2}\\
G_{LR} & = & \frac{e^{2}}{h}\sum_{\sigma,s}\int_{-\infty}^{\infty}d\omega\;\left[\gamma_{L}\gamma_{R}\frac{dn_{R}}{d\left(eV_{R}\right)}g_{1\sigma,Ns}^{r}g_{Ns,1\sigma}^{a}+\frac{d\bar{n}_{R}}{d\left(eV_{R}\right)}\gamma_{L}\bar{\gamma}_{R}f_{1\sigma,Ns}^{r}\bar{f}_{Ns,1\sigma}^{a}\right]_{\omega},\label{eq:GLR_2}\\
G_{RL} & = & \frac{e^{2}}{h}\sum_{\sigma,s}\int_{-\infty}^{\infty}d\omega\;\left[\frac{dn_{L}}{d\left(eV_{L}\right)}\gamma_{R}\gamma_{L}g_{N\sigma,1s}^{r}g_{1s,N\sigma}^{a}+\frac{d\bar{n}_{L}}{d\left(eV_{L}\right)}\gamma_{R}\bar{\gamma}_{L}f_{N\sigma,1s}^{r}\bar{f}_{1s,N\sigma}^{a}\right]_{\omega},\label{eq:GRL_2}\\
G_{RR} & = & -\frac{e^{2}}{h}\sum_{\sigma}\int_{-\infty}^{\infty}d\omega\;\gamma_{R}\left(\omega\right)\left[\frac{dn_{R}}{d\left(eV_{R}\right)}2\pi\rho_{N\sigma}-\sum_{s}\frac{dn_{R}}{d\left(eV_{R}\right)}\gamma_{R}g_{N\sigma,Ns}^{r}g_{Ns,N\sigma}^{a} \right. \nonumber \\ 
&&\left. - \sum_{s}\frac{d\bar{n}_{R}}{d\left(eV_{R}\right)}\bar{\gamma}_{R}f_{N\sigma,Ns}^{r}\bar{f}_{Ns,N\sigma}^{a}\right]_{\omega},\label{eq:GRR_2}
\end{eqnarray}
where we have defined the broadening 
\begin{eqnarray}
\gamma_{j}\left(\omega\right) & =&2\pi t_{j}^{2}\rho_{j}^{0}\left(\omega\right),\\
\bar{\gamma}_{j}\left(\omega\right) & =&2\pi t_{j}^{2}\bar{\rho}_{j}^{0}\left(\omega\right).
\end{eqnarray}
To make contact with BTK theory,\cite{blonder1982,Anantram96_Andreev_scattering}
we can express these results in a more standard form by recalling
that $M_{L}\left(\omega\right)=2\pi\text{Tr }\left[\boldsymbol{\Gamma}_{L}\left(\omega\right)\boldsymbol{\rho}_{1}\left(\omega\right)\right]=\sum_{\sigma}2\pi\gamma_{L}\left(\omega\right)\rho_{1\sigma}\left(\omega\right)$
is the number of modes in the lead $L$ at frequency $\omega$, and
$M_{R}\left(\omega\right)=2\pi\text{Tr }\left[\boldsymbol{\Gamma}_{R}\left(\omega\right)\boldsymbol{\rho}_{N}\left(\omega\right)\right]=\sum_{\sigma}2\pi\gamma_{R}\left(\omega\right)\rho_{N\sigma}\left(\omega\right)$
is the analog quantity for lead $R$, where we have defined the matrices
$\boldsymbol{\Gamma}_{L(R)}\left(\omega\right)=\left(\begin{array}{cc}
\gamma_{L(R)}\left(\omega\right) & 0\\
0 & \gamma_{L(R)}\left(\omega\right)
\end{array}\right)$, and $\boldsymbol{\rho}_{1(N)}\left(\omega\right)=2\pi\left(\begin{array}{cc}
\rho_{L(N),\uparrow}\left(\omega\right) & 0\\
0 & \rho_{L(N),\downarrow}\left(\omega\right)
\end{array}\right)$(see Ref. \cite{datta}). On the other hand, defining the matrices
\[
\begin{array}{ccl}
\mathbf{r}_{ee}^{LL}\left(\omega\right) & = & \left(\begin{array}{cc}
\gamma_{L}g_{1\uparrow,1\uparrow}^{r} & \gamma_{L}g_{1\uparrow,1\downarrow}^{r}\\
\gamma_{L}g_{1\downarrow,1\uparrow}^{r} & \gamma_{L}g_{1\downarrow,1\downarrow}^{r}
\end{array}\right)_{\omega}\\
\\
\mathbf{r}_{eh}^{LL}\left(\omega\right) & = & \left(\begin{array}{cc}
\gamma_{L}f_{1\uparrow,1\uparrow}^{r} & \gamma_{L}f_{1\uparrow,1\downarrow}^{r}\\
\gamma_{L}f_{1\downarrow,1\uparrow}^{r} & \gamma_{L}f_{1\downarrow,1\downarrow}^{r}
\end{array}\right)_{\omega}\\
\\
\mathbf{r}_{ee}^{RR}\left(\omega\right) & = & \left(\begin{array}{cc}
\gamma_{R}g_{N\uparrow,N\uparrow}^{r} & \gamma_{R}g_{N\uparrow,N\downarrow}^{r}\\
\gamma_{R}g_{N\downarrow,N\uparrow}^{r} & \gamma_{R}g_{N\downarrow,N\downarrow}^{r}
\end{array}\right)_{\omega}\\
\\
\mathbf{r}_{eh}^{RR}\left(\omega\right) & = & \left(\begin{array}{cc}
\gamma_{R}f_{N\uparrow,N\uparrow}^{r} & \gamma_{R}f_{N\uparrow,N\downarrow}^{r}\\
\gamma_{R}f_{N\downarrow,N\uparrow}^{r} & \gamma_{R}f_{N\downarrow,N\downarrow}^{r}
\end{array}\right)_{\omega}\\
\\
\mathbf{t}_{ee}^{LR}\left(\omega\right) & = & \left(\begin{array}{cc}
\sqrt{\gamma_{L}\gamma_{R}}g_{1\uparrow,N\uparrow}^{r} & \sqrt{\gamma_{L}\gamma_{R}}g_{1\uparrow,N\downarrow}^{r}\\
\sqrt{\gamma_{L}\gamma_{R}}g_{N\downarrow,N\uparrow}^{r} & \sqrt{\gamma_{L}\gamma_{R}}g_{1\downarrow,N\downarrow}^{r}
\end{array}\right)_{\omega}\\
\\
\mathbf{t}_{eh}^{LR}\left(\omega\right) & = & \left(\begin{array}{cc}
\sqrt{\gamma_{L}\gamma_{R}}f_{1\uparrow,N\uparrow}^{r} & \sqrt{\gamma_{L}\gamma_{R}}f_{1\uparrow,N\downarrow}^{r}\\
\sqrt{\gamma_{L}\gamma_{R}}f_{N\downarrow,N\uparrow}^{r} & \sqrt{\gamma_{L}\gamma_{R}}f_{1\downarrow,N\downarrow}^{r}
\end{array}\right)_{\omega}
\end{array}
\]
 we can express our Eqs. (\ref{eq:GLL_2})-(\ref{eq:GRR_2}) in the
BTK language as \cite{blonder1982,Anantram96_Andreev_scattering}

\begin{eqnarray}
G_{LL} & =&\frac{e^{2}}{h}\int_{-\infty}^{\infty}d\omega\;\left\{ -\frac{dn_{L}}{d\left(eV_{L}\right)}M_{L}+\frac{dn_{L}}{d\left(eV_{L}\right)}\text{Tr }\left[\mathbf{r}_{ee}^{LL}\left(\mathbf{r}_{ee}^{LL}\right)^{\dagger}\right]+\frac{d\bar{n}_{L}}{d\left(eV_{L}\right)}\text{Tr }\left[\mathbf{r}_{eh}^{LL}\left(\mathbf{r}_{eh}^{LL}\right)^{\dagger}\right]\right\} _{\omega},\label{eq:GLL_BTK}\\
G_{LR} & =&\frac{e^{2}}{h}\int_{-\infty}^{\infty}d\omega\;\left\{ \frac{dn_{R}}{d\left(eV_{R}\right)}\text{Tr }\left[\mathbf{t}_{ee}^{LR}\left(\mathbf{t}_{ee}^{LR}\right)^{\dagger}\right]+\frac{d\bar{n}_{R}}{d\left(eV_{R}\right)}\text{Tr }\left[\mathbf{t}_{eh}^{LR}\left(\mathbf{t}_{eh}^{LR}\right)^{\dagger}\right]\right\} _{\omega},\label{eq:GLR_BTK}\\
G_{RL} & =&\frac{e^{2}}{h}\int_{-\infty}^{\infty}d\omega\;\left\{ \frac{dn_{L}}{d\left(eV_{L}\right)}\text{Tr }\left[\mathbf{t}_{ee}^{RL}\left(\mathbf{t}_{ee}^{RL}\right)^{\dagger}\right]+\frac{d\bar{n}_{L}}{d\left(eV_{L}\right)}\text{Tr }\left[\mathbf{t}_{eh}^{RL}\left(\mathbf{t}_{eh}^{RL}\right)^{\dagger}\right]\right\} _{\omega},\label{eq:GRL_BTK}\\
G_{RR} & =&\frac{e^{2}}{h}\int_{-\infty}^{\infty}d\omega\;\left\{ -\frac{dn_{R}}{d\left(eV_{R}\right)}M_{R}+\frac{dn_{R}}{d\left(eV_{R}\right)}\text{Tr }\left[\mathbf{r}_{ee}^{RR}\left(\mathbf{r}_{ee}^{RR}\right)^{\dagger}\right]+\frac{d\bar{n}_{R}}{d\left(eV_{R}\right)}\text{Tr }\left[\mathbf{r}_{eh}^{RR}\left(\mathbf{r}_{eh}^{RR}\right)^{\dagger}\right]\right\} _{\omega},\label{eq:GRR_BTK}
\end{eqnarray}
where for convenience we have omitted the argument $\omega$ inside
the brackets. 

In order to make explicit the non-local terms in these expressions
we make use of the identity \cite{datta} 
\begin{eqnarray}
\boldsymbol{\mathcal{G}}^{r}\left(\omega\right)-\boldsymbol{\mathcal{G}}^{a}\left(\omega\right) & =&\boldsymbol{\mathcal{G}}^{r}\left(\omega\right)\left[\boldsymbol{\Sigma}^{r}\left(\omega\right)-\boldsymbol{\Sigma}^{a}\left(\omega\right)\right]\boldsymbol{\mathcal{G}}^{a}\left(\omega\right),
\end{eqnarray}
 From here, the following results are obtained

\begin{eqnarray}
g_{1\sigma,1\sigma}^{r}\left(\omega\right)-g_{1\sigma,1\sigma}^{a}\left(\omega\right) & =&-2\pi i\rho_{1,\sigma}\left(\omega\right)=-2\pi i\sum_{s} \left[t_{L}^{2}\rho_{L}^{0}g_{1\sigma,1s}^{r}g_{1s,1\sigma}^{a}\right. \nonumber\\
 & &\left.+t_{L}^{2}\bar{\rho}_{L}^{0}f_{1\sigma,1s}^{r}\bar{f}_{1s,1\sigma}^{a}+t_{R}^{2}\rho_{R}^{0}g_{1\sigma,Ns}^{r}g_{Ns,1\sigma}^{a}+t_{R}^{2}\bar{\rho}_{R}^{0}f_{1\sigma,Ns}^{r}\bar{f}_{Ns,1\sigma}^{a}\right],\\
g_{N\sigma,N\sigma}^{r}\left(\omega\right)-g_{N\sigma,N\sigma}^{a}\left(\omega\right) & =&-2\pi i\rho_{N,\sigma}\left(\omega\right) =-2\pi i\sum_{s}\left[t_{R}^{2}\rho_{R}^{0}g_{N\sigma,Ns}^{r}g_{Ns,N\sigma}^{a}\right. \nonumber\\
 &&\left.+t_{R}^{2}\bar{\rho}_{R}^{0}f_{N\sigma,Ns}^{r}\bar{f}_{Ns,N\sigma}^{a}+t_{L}^{2}\rho_{L}^{0}g_{N\sigma,1s}^{r}g_{1s,N\sigma}^{a}+t_{L}^{2}\bar{\rho}_{L}^{0}f_{N\sigma,1s}^{r}\bar{f}_{1s,N\sigma}^{a}\right], 
\end{eqnarray}
 and hence, substituting into Eqs. (\ref{eq:GLL_BTK})-(\ref{eq:GRR_BTK}),
we obtain

\begin{eqnarray}
G_{LL} & =&\frac{e^{2}}{h}\int_{-\infty}^{\infty}d\omega\;\left[-\frac{dn_{L}\left(\omega\right)}{d\left(eV_{L}\right)}\right]\biggl\{2\text{Tr }\left[\mathbf{r}_{eh}^{LL}\left(\mathbf{r}_{eh}^{LL}\right)^{\dagger}\right]+\text{Tr }\left[\mathbf{t}_{ee}^{LR}\left(\mathbf{t}_{ee}^{LR}\right)^{\dagger}\right]+\text{Tr }\left[\mathbf{t}_{eh}^{LR}\left(\mathbf{t}_{eh}^{LR}\right)^{\dagger}\right]\biggr\}_{\omega},\label{eq:GLL_final_symm}\\
G_{LR} & =&\frac{e^{2}}{h}\int_{-\infty}^{\infty}d\omega\;\left[\frac{dn_{R}\left(\omega\right)}{d\left(eV_{R}\right)}\right]\biggl\{\text{Tr }\left[\mathbf{t}_{ee}^{LR}\left(\mathbf{t}_{ee}^{LR}\right)^{\dagger}\right]-\text{Tr }\left[\mathbf{t}_{eh}^{LR}\left(\mathbf{t}_{eh}^{LR}\right)^{\dagger}\right]\biggr\}_{\omega},\label{eq:GLR_final_symm}\\
G_{RL} & =&\frac{e^{2}}{h}\int_{-\infty}^{\infty}d\omega\;\left[\frac{dn_{L}\left(\omega\right)}{d\left(eV_{L}\right)}\right]\biggl\{\text{Tr }\left[\mathbf{t}_{ee}^{RL}\left(\mathbf{t}_{ee}^{RL}\right)^{\dagger}\right]-\text{Tr }\left[\mathbf{t}_{eh}^{RL}\left(\mathbf{t}_{eh}^{RL}\right)^{\dagger}\right]\biggr\}_{\omega},\label{eq:GRL_final_symm}\\
G_{RR} & =&\frac{e^{2}}{h}\int_{-\infty}^{\infty}d\omega\;\left[-\frac{dn_{R}\left(\omega\right)}{d\left(eV_{R}\right)}\right]\biggl\{2\text{Tr }\left[\mathbf{r}_{eh}^{RR}\left(\mathbf{r}_{eh}^{RR}\right)^{\dagger}\right]+\text{Tr }\left[\mathbf{t}_{ee}^{RL}\left(\mathbf{t}_{ee}^{RL}\right)^{\dagger}\right]+\text{Tr }\left[\mathbf{t}_{eh}^{RL}\left(\mathbf{t}_{eh}^{RL}\right)^{\dagger}\right]\biggr\}_{\omega},\label{eq:GRR_final_symm}
\end{eqnarray}

which correspond to Eqs. (\ref{eq:GLL_final_symm-1})-(\ref{eq:GRR_final_symm-1}).
\bibliographystyle{unsrtnat}

\begin{thebibliography}{77}
\bibitem[not()]{note_majorana_long_paper}
The oxide layer that inevitably forms around the semiconductor nanowires needs
  to be removed in order to create a good contact with the superconductor, a
  step which is typically done by ammonium sulfide etching, sometimes followed
  by Ar plasma etching. The amophous superconducting layer is then deposited on
  top of the semiconductor by sputtering techniques. While this process is
  effective in removing the oxide layer in the semiconductor, it inevitably
  damages the nanowire surface.
\end{thebibliography}

\end{document}